\documentclass{pasj01}
\usepackage[switch,mathlines]{lineno}
\Received{2024 July 1}
\Accepted{2024 November 26}
\Published{$\langle$2024 December 26$\rangle$}

\usepackage{setspace}

\begin{document}
\title{The possible long-term periodic variability of the extremely luminous quasar WISE J090924.01+000211.1}
\author{Takashi H{\small ORIUCHI},$^{1,*}$ Yoshiki T{\small OBA},$^{2,3,4}$ Toru M{\small ISAWA}$^{5}$, 
Katsuhiro L. M{\small URATA},$^{6}$ Keisuke I{\small SOGAI},$^{6,7}$  Yoichi Y{\small ATSU},$^{8}$ 
Ichiro T{\small AKAHASHI},$^{8}$ Mahito S{\small ASADA},$^{9}$ 
Masafumi N{\small IWANO},$^{8}$ Narikazu H{\small IGUCHI},$^{8}$ Shunsuke H{\small AYATSU},$^{8}$
Hibiki S{\small EKI},$^{8}$ Yumiko O{\small ASA},$^{10,11,12}$ and Rikuto S{\small ATO}$^{11}$}%
\altaffiltext{1}{Institute of Astronomy, Graduate School of Science, The University of 
Tokyo, 2-21-1, Osawa, Mitaka, Tokyo 181-0015, Japan}
\altaffiltext{2}{National Astronomical Observatory of Japan, 2-21-1 Osawa, Mitaka, Tokyo 181-8588, Japan}
\altaffiltext{3}{Academia Sinica Institute of Astronomy and Astrophysics, 11F of Astronomy-Mathematics Building, AS/NTU, No.1, Section 4, Roosevelt Road, Taipei 10617, Taiwan}
\altaffiltext{4}{Research Center for Space and Cosmic Evolution, Ehime University, 2-5 Bunkyo-cho, Matsuyama, Ehime 790-8577, Japan}
\altaffiltext{5}{Center for General Education, Shinshu University, 3-1-1 Asahi, Matsumoto, Nagano 390-8621, Japan}
\altaffiltext{6}{Okayama Observatory, Kyoto University, 3037-5 Honjo, Kamogata-cho, Asakuchi, Okayama 719-0232, Japan}
\altaffiltext{7}{Department of Multi-Disciplinary Sciences, Graduate School of Arts and Sciences, The University of Tokyo, 3-8-1 Komaba, Meguro, Tokyo 153-8902, Japan}
\altaffiltext{8}{Department of Physics, Institute of Science Tokyo, 2-12-1 Ookayama, Meguro-ku, Tokyo 152-8551, Japan}
\altaffiltext{9}{Institute of Integrated Research, Institute of Science Tokyo, 2-12-1 Ookayama, Meguro-ku, Tokyo 152-8550, Japan}
\altaffiltext{10}{Graduate School of Science and Engineering, Saitama University, 255 Shimo-Okubo, Sakura-ku, Saitama, Saitama 338-8570, Japan}
\altaffiltext{11}{Graduate School of Education, Saitama University, 255 Shimo-Okubo, Sakura-ku, Saitama, Saitama 338-8570, Japan}
\altaffiltext{12}{Faculty of Education, Saitama University, 255 Shimo-Okubo, Sakura-ku, Saitama, Saitama 338-8570, Japan}
\email{t-horiuchi@ioa.s.u-tokyo.ac.jp}

\maketitle

\begin{abstract}
The extremely luminous infrared galaxy (ELIRG), WISE J090924.01+000211.1 (hereafter; WISE J0909+0002, $z=1.87$) is an extraordinary object with a quasar aspect. This study performs monitoring observations of WISE J0909+0002 with the 105 cm Murikabushi telescope, Okayama and Akeno 50 cm telescopes/MITSuME ($g'$, $R_{\rm c}$, and $I_{\rm c}$ bands), and the SaCRA 55 cm telescope/MuSaSHI ($r$, $i$, and $z$ bands). We obtain the following results by combining the UV/optical light curves of the CRTS, Pan-STARRS, and ZTF archive data, and our observational data: (1) the light curves of WISE J0909+0002 present quasi-periodic (sinusoidal) oscillations with the rest-frame period of $\sim$ 660$-$689 day; (2) the structure functions of WISE J0909+0002 do not show a damped random walk (DRW) trend; (3) the mock DRW light curves present periodic-like trend on rare occasions in 10000 simulations; (4) the relativistic boost scenario is favored, since the relation between variability amplitude and power-law slope ratio is consistent with the theoretical prediction of this scenario, and a substantial parameter space exists between the inclination angles and the black hole mass; (5) the circumbinary disk model is difficult to explain the spectral energy distribution of our target; (6) the significant radio flux density of WISE J0909+0002 is not detected from the VLA FIRST Survey, thus the radio jet precession scenario is ruled out. From our results, the Doppler boost scenario is likely as a cause of the periodic variability, consequently the quasi-periodic oscillations in WISE J0909+0002 is possibly interpreted by a supermassive blackhole binary. Additional observations to investigate the continuity of the periodic trend would bring new insights into mechanisms of the quasi-periodic oscillations and/or ELIRGs.
\end{abstract}


\section{Introduction}
\subsection{Periodic variability of active galactic nuclei}
The flux variability of active galactic nuclei (AGNs) is a universal phenomenon 
from X-ray to radio wavelengths. The AGN variability have often been described 
with the damped random walk (DRW) model (e.g., \cite{2009ApJ...698..895K}; 
\cite{2010ApJ...708..927K}; \cite{2010ApJ...721.1014M}). 
However, the periodic (sinusoidal) flux variability with a period of 1884 day was 
discovered in the quasar PG 1302$-$102 \citep{2015...518..74G}. 

Such quasi-periodic oscillations (QPOs) have been actively studied, for instance, 
the Pan-STARRS1 Medium Deep Survey data with $griz$ bands identified 26 candidates 
of periodic light curves \citep{2019ApJ...884...36L}. A recent study identified 106 (DRW as 
a null hypothesis) and 86 (single power-law as a null hypothesis) candidates of periodic 
quasars from the photometric data of the Zwicky Transient Facility \citep{2024MNRAS.52712154C}. 
Overall, the detection rate of quasars with QPOs is $\sim$ 0.01 $-$ 1.1$\%$ 
(\cite{2016MNRAS.463.2145C}; \cite{2019ApJ...884...36L}; \cite{2020MNRAS.499.2245C}) 
As examples for individual objects, half to several-year periodicity was discovered in 
the following AGNs: SDSS J025214.67-002813.7 ($z=1.53$; \cite{2021MNRAS.500.4025L}), 
Q J0158-4325 ($z=1.29$; \cite{2022A&A...668A..77M}), PKS 2131-021 ($z=1.285$; 
\cite{2022ApJ...926L..35O}), SDSS J132144+033055 ($z=0.269$; \cite{2022MNRAS.516.3650Z}), 
SDSS J143016.05+230344.4 ($z=0.08$; \cite{2022A&A...665L...3D}; 
\cite{2024PASJ...76..103H}). The light curve modeling for SDSS 
J025214.67-002813.7 was improved by considering periodic components 
\citep{2022MNRAS.513.2841C}. 

While the mechanism of the periodic light curves is not well known, a binary 
supermassive blackhole (BSBH) within a mili to sub-parsec scale is the most 
likely candidate of that phenomenon (e.g., \cite{2006MmSAI..77..733K}; 
\cite{2015...518..74G}; \cite{2015Natur.525..351D}; \cite{2020ApJ...901...25D}). 
That is to say, AGNs with periodic light curves are possibly the site emitting gravitational 
waves (e.g., \cite{2023ApJ...951L..50A}). The BSBH system is thought to yield 
an orbital motion of primary and secondary mini-accretion disks with a relativistic 
Doppler boost (\cite{2015Natur.525..351D}). The Doppler boost 
scenario is partially supported by previous studies (e.g., \cite{2018MNRAS.476.4617C}; 
\cite{2020MNRAS.496.1683X}). However, the All-sky Automated Survey for 
Supernovae (ASAS-SN) light curve of PG 1302-102 presented a perturbation 
in its periodicity, which was identified as a false positive signal and do not 
support the BHSH theory (\cite{2018ApJ...859L..12L}; 
\cite{2019ApJ...871...32K}).  

Other plausible scenarios of the periodic light curves, for instance, are 
accretion-disc precession (e.g., \cite{1995ApJ...438..610E}; \cite{2003ApJ...598..956S}), 
and microlensing by binary stars. The former scenario was denied by the 
analysis of \citet{2022MNRAS.516.3650Z}. The later was also ruled out in the lensed quasar 
Q J0158-4325, since this scenario required a $\sim2000$-yr period from dynamical simulation 
of a binary star with the microlensing \citep{2022A&A...668A..77M}. 
At present there is no clear consensus on the mechanism of QPOs in AGNs. 
In order to overcome this situation, we focused on extremely luminous infrared 
galaxies since they are believed to be in the final stage of galaxy (and may be SMBH) 
merging (ELIRGs; \cite{2015ApJ...805..90Z}; \cite{2018ApJ...857...31T}, 2020).

\subsection{ELIRGs and our target} 
The galaxy mergers are important phases of the galaxy evolution. 
It is known that ultra-luminous infrared (IR) galaxies (ULIRGs) 
and Hyper-luminous IR galaxies (HyLIRGs; \cite{2000MNRAS316..815R}) 
indicate the evidence of the final stage of galaxy interactions. Within the last 
decade, a series of ELIRGs with the infrared luminosity of $>$ 10$^{14}$ $L_\odot$ 
are discovered, and these objects also exhibit a sign of galaxy interaction with 
thick dust. Since previous studies consider that such infrared-luminous galaxies 
are the peak stage of the coevolution of galaxies (e.g., \cite{2008ApJS..175..356H}; 
\cite{2018MNRAS.478.3056B}; \cite{2022ApJ...936..118Y}), ELIRGs are important 
research targets. \citet{2021A&A...649L..11T} newly identified WISE 
J090924.01+000211.1 (RA: \timeform{09h09m24s}, Dec: \timeform{00D02'11"}, $z=1.87$; 
hereafter WISE J0909+0002) as an ELIRG and reported its interesting properties: 
(1) the infrared luminosity, $L_{\rm IR}~=$ 1.79 $\times$ 10$^{14}$ $L_\odot$, (2) 
the SDSS spectrum presents broad emission lines, and the estimated blackhole mass 
is extremely massive with a mass of 7.4 $\times$ 10$^9$ $M_{\odot}$, 
and (3) the extremely high infrared luminosity with quite low column density may 
imply that this object is in the midst of a short-lived phase. As shown in figure 1, 
WISE J0909+0002 presents an exceedingly high bolometric luminosity compared to 
most quasars. Namely, the ELIRG, WISE J0909+0002 includes an extremely 
luminous type1 quasar. However, the internal structure of WISE J0909+0002 is still 
unclear due to its outlandish properties. 

We independently found the periodic variability trend of WISE J0909+0002 from 
archival data and our monitoring observations. This study provides QPOs 
of this object and the related analysis of its light curves. We describe the 
details of photometric data of WISE J0909+0002 from archives and observation 
with the observation collaboration of optical and infrared synergetic telescopes 
for education and research (OISTER) in section 2. 
Section 3 presents the flux variability trend WISE J0909+0002, and we then 
discuss the inferred physical properties of this object in section 4. The conclusion 
of our study is summarized in section 5. A ${\rm \Lambda}$CDM cosmology 
with $H_{0}=~$70 km s$^{-1}$ Mpc$^{-1}$, $\Omega_{m}$~=~0.3, and 
$\Omega_{\Lambda}$~=~0.7 is adopted in this study. 

\begin{figure}
 \begin{center}
   \includegraphics[height=6.5cm,width=8cm]{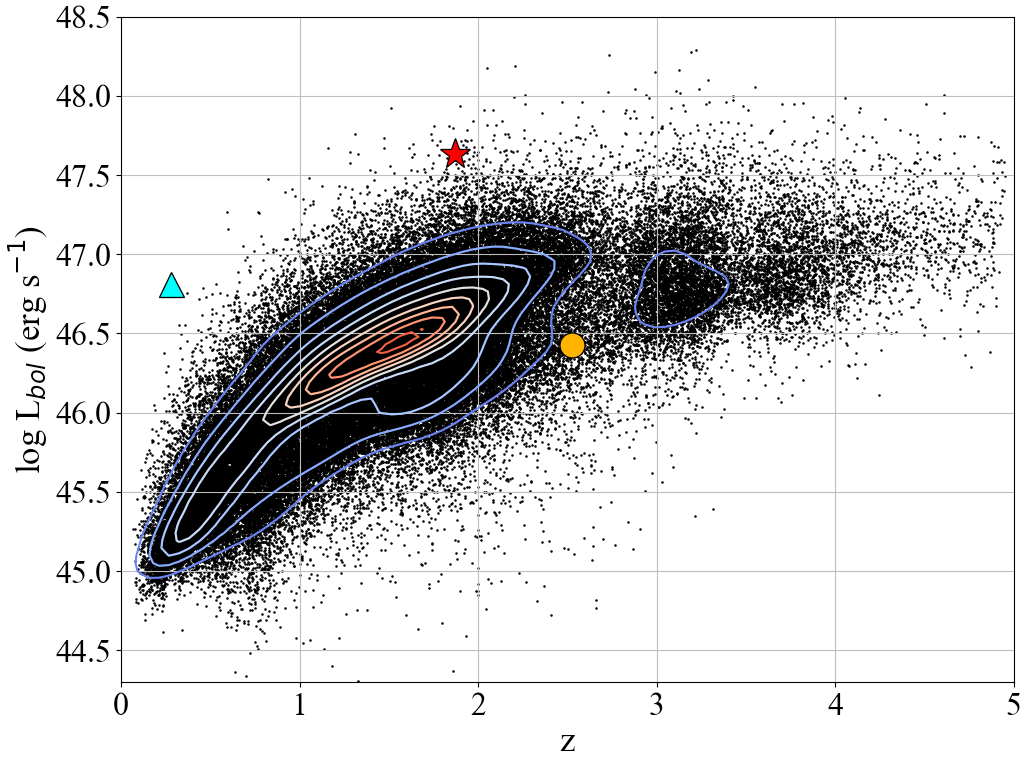} 
 \end{center}
\caption{The redshift dependence of quasar bolometric luminosity from 
the Sloan Digital Sky Survey (SDSS) Data Release 7 \citep{2011ApJS...194..45S}. 
The redder (or bluer) contours indicate the denser (or thiner) area of 
this distribution. The red star, orange circle, and cyan triangle represent the 
data points for WISE J0909+0002 from \citet{2021A&A...649L..11T}, the periodic 
quasars J024703.24-010032.0 from \citet{2020MNRAS.499.2245C}, and PG 1302-102, 
respectively.}
\end{figure}

\section{Archive data and our observation}
\subsection{Archive data} 
In order to measure the flux variability of WISE J0909+0002, we use the archival 
photometric data as follows: the Catalina Real-Time Transient Survey (CRTS; 
\cite{2009ApJ...696..870D}; \cite{Mahabal2011}), the Panoramic Survey 
Telescope and Rapid Response System (Pan-STARRS; \cite{2016arXiv161205560C}), 
and Zwicky Transient Facility (ZTF). In addition, we conducted the follow-up observation 
of WISE J0909+0002 with the OISTER collaboration. Table 1 summarizes details of 
those archival and observation data. Details of those photometric 
data are explained in the following sub-subsections. 

\subsubsection{CRTS data} 
This survey operates three telescopes: Catalina Sky Survey (CSS) 0.7 m 
Schmidt, Mount Lemmon Survey (MLS) 1.5 m Cass in Arizona, and 
Siding Springs Survey (SSS) 0.5 m Schmidt in Australia. Those telescopes 
sweep sky fields up to $\sim$ 2,500 deg$^2$ with four times observations 
per visit on a clear night. The archival data taken by CRTS are converted 
to the $V$ band magnitude based on 50-100 G-type stars using 2MASS data 
\citep{2013ApJ...763...32D}. The CSS light curve of WISE J0909+0002 
covers the period from MJD 53494 to 56397 with 330 data points. During 
this period, the light curve had showed a QPO trend (see subsection 3.1). 

\subsubsection{Pan-STARRS data} 
The Pan-STARRS carries out three-day cadence $grizy$ band observations 
with the 1.8 m telescope. The $grizy$ bandwidths are listed in \citet{2012ApJ...750...99T}. 
We converted “psfFlux” (in unit of Jy) into an AB magnitude of which the zero-point 
flux is 3631 Jy. The epoch of the photometric data for WISE J0909+0002 ranges 
from MJD 55239.4 to 57021.5.

\subsubsection{ZTF data} 
The ZTF survey scans the entire Northern sky with 2-day cadence observations, 
using the 48-inch Samuel Oschin Schmidt telescope with the ZTF $g$, $r$, and $i$ 
filters (e.g., \cite{2019PASP..131a8002B}). We obtained 5 (for ZTF $g$) and 42 
(for ZTF $r$) photometric data without bad-data flags (i.e., {\tt catflags}$~=~0$) 
from the ZTF Data Release 20. The epoch of those data covers the observation period 
from MJD 58430.5 to 59271 (or MJD 58439.4 to 59638.2) for the ZTF $g$ (or $r$) band. 
The ZTF magnitudes were converted to those of the Pan-STARRS by 
\citet{2020RNAAS...4...38M}. 

\subsection{Monitoring observations}
From 2021, February to 2022, March, we performed $g'$ (4770 ${\rm \AA}$), 
$R_{\rm c}$ (6492 ${\rm \AA}$), and $I_{\rm c}$ (8020 ${\rm \AA}$) bands follow-up 
observations ($\sim$ 1 day to a week cadence) of WISE J0909+0002 with the 
105 cm Murikabushi telescope/MITSuME. From 2023, May to 2024, April we had 
observed the same object with the Okayama and Akeno 50 cm telescopes/MITSuME 
with the $g'$, $R_{\rm c}$, and $I_{\rm c}$ bands (\cite{2005NCimC..28..755K}; 
\cite{2007PhyE...40..434Y}; \cite{2008AIPC.1000..543S}) and SaCRA 55 cm 
telescope/MuSaSHI \citep{2020SPIE11447E..5ZO} with  $r$, $i$, and $z$ bands 
as the OISTER collaboration to confirm the continuous periodicity flux variability of 
our target. The magnitudes of WISE J0909+0002 obtained from those telescopes and 
instruments were converted to the AB magnitude based on the bandpass transformations 
of \citet{2012ApJ...750...99T} and the relation between the Vega and AB magnitude 
\citep{2007AJ....133..734B}. We used photometric data of the signal-to-noise ratio of 
$\gtrsim 10$ taken with those telescopes and removed those data with a wrong 
weather condition and astronomical seeing. 

\begin{table*}
\tbl{The list of archival and our observation data of WISE J0909+0002.}{
\begin{tabular}{cccccc}
\hline
\multicolumn{6}{c}{Archive data} \\
\hline
 & Filter & $N$\footnotemark[$*$] & $m$\footnotemark[$\dagger$] & $\sigma_{m}$\footnotemark[$\ddagger$] & Observation epoch \\
&  & & (mag) & (10$^{-2}$ mag) & (MJD) \\  
\hline
CRTS  & $V$ & 330 & 16.46 & 7.0 & 53494.1 $-$ 56397.1 \\ 
Pan-STARRS & $g$ & 16 & 16.83  & 0.5 & 55239.4 $-$ 56715.4 \\ 
& $r$ & 7 & 16.72 & 0.5  &  56315.4 $-$ 56662.6 \\ 
& $i$ & 21 & 16.32 & 0.4 &55577.4 $-$ 57021.5 \\ 
& $z$ & 15 & 16.39 & 0.7 & 55284.3 $-$ 56991.6 \\ 
& $y$ & 15 & 16.37 & 1.0 & 55523.5 $-$ 56991.6 \\ 
ZTF& $g$ & 5 & 16.72 & 1.0 & 58430.5 $-$ 59271.3\\ 
& $r$ & 42 & 16.59 & 1.0 & 58439.3 $-$ 59638.2 \\ 
\hline
 \multicolumn{6}{c}{Our observations} \\
\hline
Murikabushi/MITSuME & $g'$ & 35 & 16.75 & 1.0 & 59257.6 $-$ 59619.6 \\ 
  & $R_{\rm c}$ & 35 & 16.68 & 0.9 & 59257.6 $-$ 59619.6 \\ 
  & $I_{\rm c}$ & 34 & 16.38 & 1.0 & 59257.6 $-$ 59619.6\\ 
Akeno and Okayama/MITSuME & $g'$ & 33  & 16.70 & 3.0 & 60079.5 $-$ 60401.5\\ 
  & $R_{\rm c}$ & 42 & 16.60 & 3.0 & 60079.5 $-$ 60401.5\\ 
  & $I_{\rm c}$ & 33 & 16.45 & 4.0 &60079.5 $-$  60401.5\\ 
SaCRA/MuSaSHI & $r$ & 4 & 16.60 & 3.0 & 60269.7 $-$ 60419.5 \\ 
  & $i$ & 4 & 16.42 & 2.0 & 60269.7 $-$ 60419.5 \\ 
  & $z$ & 4 & 16.36 & 2.0 & 60269.7 $-$ 60419.5 \\ 
\hline\noalign{\vskip3pt} 
\end{tabular}}\label{table:extramath}
\begin{tabnote}
\hangindent6pt\noindent
\hbox to6pt{\footnotemark[$*$]\hss}\unskip
Number of data points.\\
\hbox to6pt{\footnotemark[$\dagger$]\hss}\unskip
Median magnitudes over the observation period.\\
\hbox to6pt{\footnotemark[$\ddagger$]\hss}\unskip
Median photometric errors over the observation period.\\
\end{tabnote}
\end{table*}

\section{Data analysis and results}

\subsection{Light curves of WISE J0909+0002}
First, we applied sinusoidal models to the light curves of WISE J0909+0002 as 
follows: 

\begin{eqnarray}
F(t) = A\sin\biggr(\frac{2\pi t}{P_{\rm Fit}}~+~\phi\biggr) + b, 
\end{eqnarray}
where $A$, $P_{\rm Fit}$, $\phi$, and $b$ are the best-fit amplitude, period, phase, and 
average magnitude, respectively. The Python package, $\tt{curve\_fit}$, was used for 
calculations of these parameters in each band. The $y$- and $R_{\rm c}$-band 
parameters were omitted from the best-fit results, since the amplitude of these two 
bands were consistent with those errors (i.e., $A~<~2\sigma$ uncertainty). 
Second, we combined all band light curves into one, since our light curves include a 
sparse period from MJD 57021 to MJD 58430. In order to connect those light curves, 
the best-fit amplitudes and median values of each band were adjusted with 
respect to the CRTS $V$-band data; hereafter we define this curve as the combined 
light curve. Finally, we calculated the signal-to-noise ratio (SNR), $\xi=A^2/(2\sigma_{\rm res}^2)$, 
of our light curves, where $\sigma_{\rm res}^2$ is the variance of the residues between 
the sinusoidal curve model and each data point \citep{1986ApJ...302..757H}. If $\xi>0.5$, 
the periodical signal is significantly larger than the residual noise (see \cite{2020MNRAS.499.2245C}). 

Figures 2 and 3 show each band light curve and combined light curve, respectively.
The $z$- and $I_{\rm c}$-band light curves exhibited very different trends from the other 
curves. We note that the difference in the phase $\phi$ in each band is not real value but just 
fitting results, since there is the disparity in observation epochs and cadences between 
each band. Table 2 lists those parameters of the all band light curves and combined light 
curve. The SNRs indicated significant periodic signals in all cases (i.e., $\xi>0.5$).

\begin{figure*}
 \begin{center}
   \includegraphics[height=12cm,width=15.5cm]{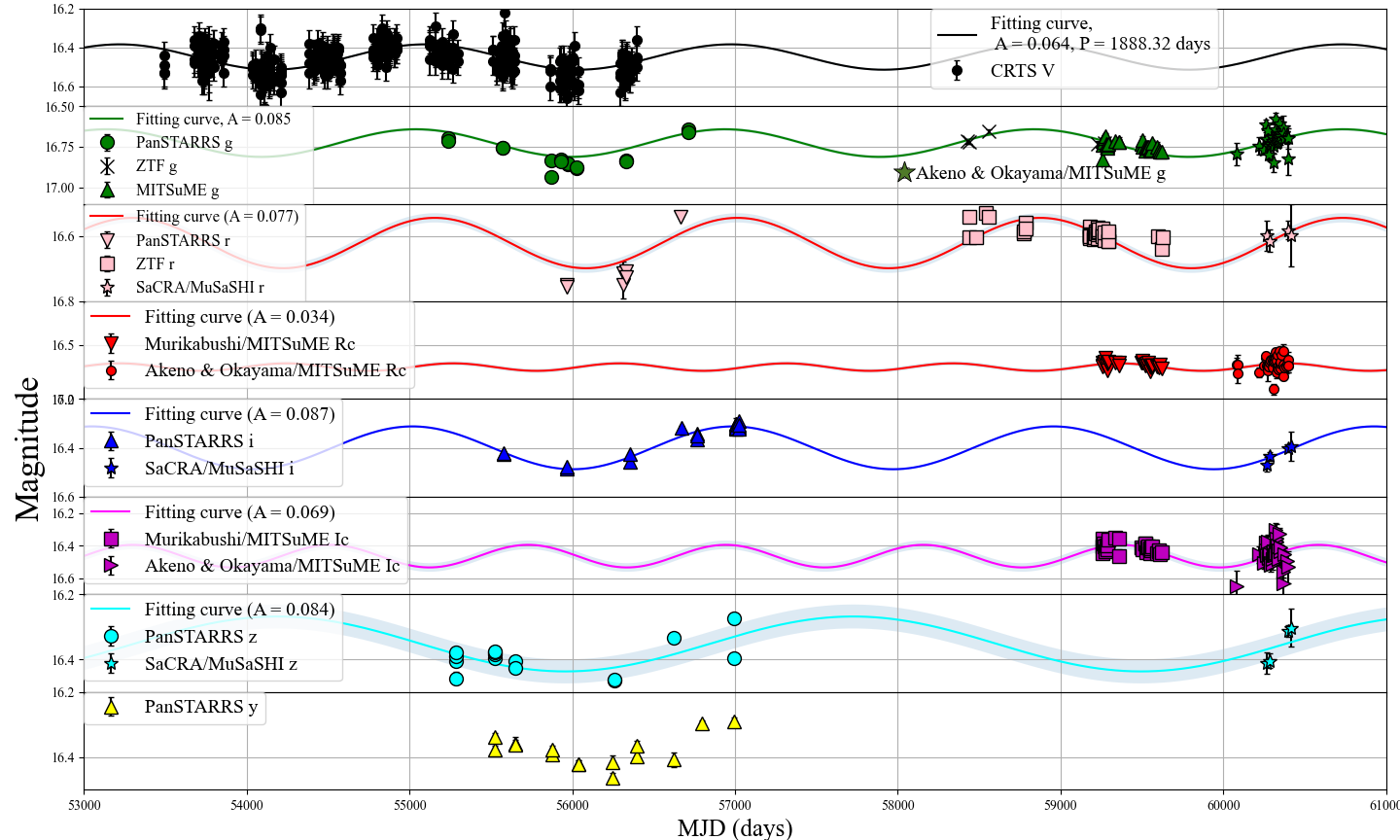} 
 \end{center}
\caption{The light curves of WISE J0909+0002 with the CRTS $V$, 
Pan-STARRS ($grizy$), ZTF ($g$ and $r$), Murikabushi telescope, Akeno 
and Okayama telescope/MITSuME ($g$, $R_{\rm c}$, and $I_{\rm c}$), and 
SaCRA/MuSaSHI ($r$, $i$, and $z$). The solid sinusoidal curves 
present the best-fitting results for these light curves. Gray shadowed region 
indicate 1$\sigma$ errors in the variability amplitude of the light curves. The $y$- 
and $R_{\rm c}$ band fitting results are not displayed, since their parameters are 
statistically insufficient.}
\end{figure*}

\begin{figure*}
 \begin{center}
   \includegraphics[height=11cm,width=15.5cm]{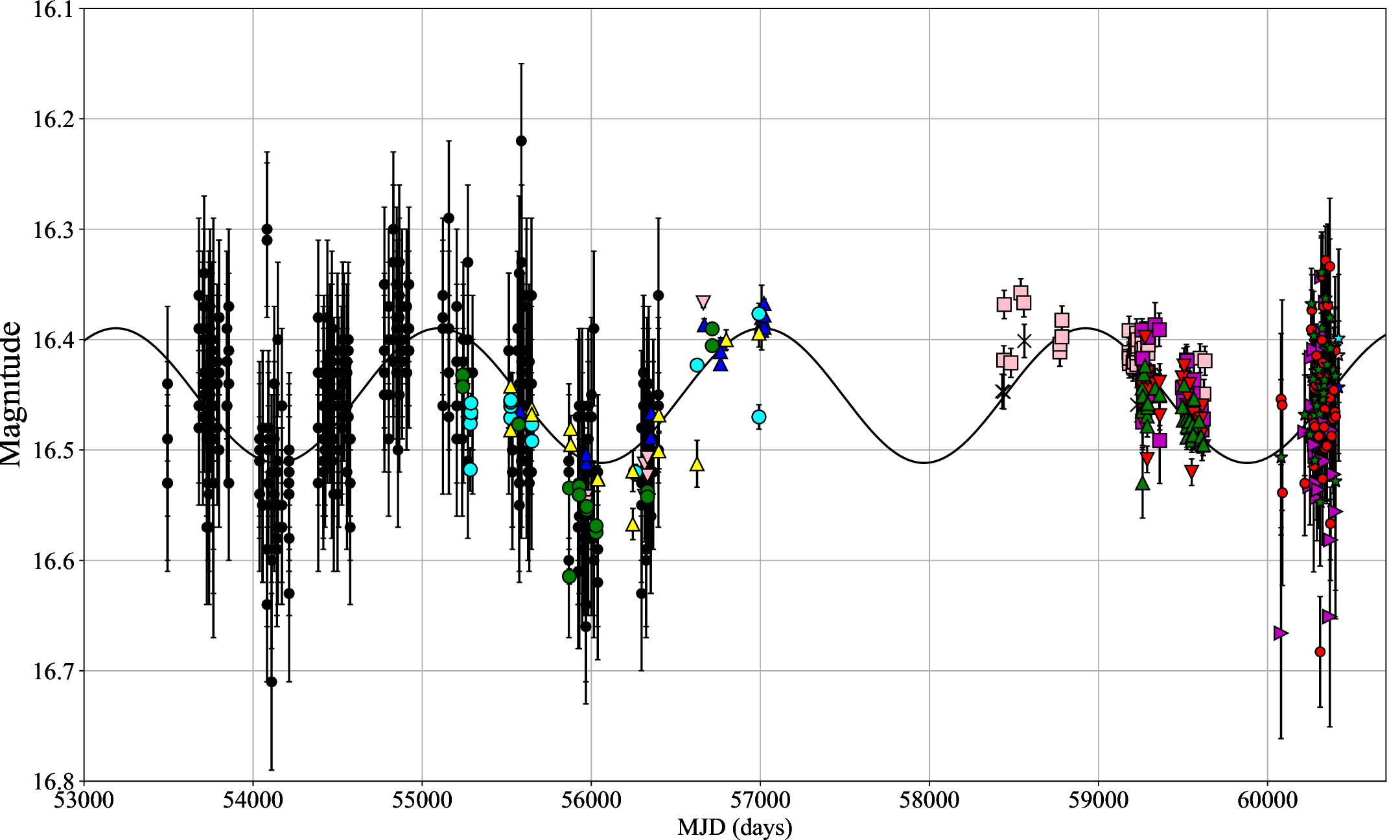} 
 \end{center}
\caption{The combined light curve of WISE J0909+0002. The symbols in this figure 
are the same as those in figure 2.}
\end{figure*}

\begin{table*}
\tbl{The sinusoidal-model parameters for the light curves.}{
\begin{tabular}{cccccc}
\hline
Band & $A$\footnotemark[$*$] & $P_{\rm Fit}$\footnotemark[$\dagger$] & $\phi$\footnotemark[$\ddagger$] & $b$\footnotemark[$\S$] & $\xi$\footnotemark[$\|$]\\
 & (mag) & (day) &  & (mag) & \\ 
\hline
CRTS $V$ & 0.064 $\pm$ 0.005 & 654.09 $\pm$ 15.24 & $-$119.97 $\pm$ 4.26 & 16.449 $\pm$ 0.003 & 0.59 \\ 
$g$ & 0.085 $\pm$ 0.011 & 661.80 $\pm$ 12.80 &  $-$117.65 $\pm$ 3.77 & 16.727 $\pm$ 0.006 & 1.62 \\ 
$r$ & 0.077 $\pm$ 0.010 & 648.09 $\pm$ 16.31 & $-$121.91 $\pm$ 4.99 & 16.620 $\pm$ 0.006 & 2.76 \\ 
$i$ & 0.087 $\pm$ 0.005 & 686.28 $\pm$ 7.10 & $-$111.10 $\pm$ 1.90 & 16.400 $\pm$ 0.004 & 18.66 \\ 
$I_{\rm c}$ & 0.067 $\pm$ 0.046 & 378.56 $\pm$ 40.58 & $-$165.89 $\pm$ 41.44 & 16.464 $\pm$ 0.034 & 0.84 \\
$z$ & 0.084 $\pm$ 0.036 & 1232.69 $\pm$ 74.26 & $-$31.82 $\pm$ 6.05 & 16.352 $\pm$ 0.024 & 2.28 \\ 
combined light curve & 0.061 $\pm$ 0.004 & 666.03 $\pm$ 3.24 & $-$116.71 $\pm$ 0.93 & 16.453 $\pm$ 0.002 & 0.68 \\ 
\hline\noalign{\vskip3pt} 
\end{tabular}}\label{table:extramath}
\begin{tabnote}
\hangindent6pt\noindent
\hbox to6pt{\footnotemark[$*$]\hss}\unskip
Variability amplitude.\\
\hbox to6pt{\footnotemark[$\dagger$]\hss}\unskip
Rest-frame periodicity.\\
\hbox to6pt{\footnotemark[$\ddagger$]\hss}\unskip
Phase.\\
\hbox to6pt{\footnotemark[$\S$]\hss}\unskip
Average magnitude.\\
\hbox to6pt{\footnotemark[$\|$]\hss}\unskip
SNR, $A^2/(2\sigma_{\rm res}^2)$.\\
\end{tabnote}
\end{table*}

\subsection{Lomb-Scargle periodgram}
To evaluate the rest-frame periods of the combined and each band light curve, 
we calculated power spectrum of the archival and our light curves by the Lomb-Scargle 
method (\cite{1976ApSS...39..447L}; \cite{1982ApJ...263..835S}; \cite{2018ApJS...236..16V}) 
with its corresponding package, $\tt{LombScargle}$ from $\tt{astropy}$. 
This method has the advantage of being able to process signals sampled 
at indefinite intervals and with missing samples. Whether the obtained power 
peaks are pure noise or not was determined by estimating the false alarm 
probability (FAP) with the bootstrap method. In this study, the periodicity is 
significant if the following conditions are satisfied: (i) the FAP at a power peak 
is significantly small ($<1.0\%$), (ii) the monitoring length of each band data 
are longer than their periods, and (iii) the time interval without observations is 
shorter than one period of the light curve. As a result the CRTS $V$-, $g$-, 
and $r$-band light curves (hereafter, three-band light curves) satisfied the 
criteria (i), (ii), and (iii). 

Figure 4 presents the Lomb-Scargle power of the combined light curve and 
three-band light curves. Despite different observation epochs of the three-band light 
curves, their rest-frame periods ($\sim$ 660$-$689 day) were consistent with 
each other. The FAPs at the peak of the combined and three-band light curves 
($<$ 0.001 for all cases) indicate that those light curves are significantly periodic. 
Therefore the light curves of WISE J0909+0002 obtained from MJD 53494.1 (CRTS $V$) 
to MJD 60419.5 (SaCRA/MuSaSHI $r$, $i$, and $z$) had likely continued the periodic 
variability at least for $\sim$ 6.6 yr in the quasar rest frame, corresponding to 
$\sim$ 3.6 cycles. 

\begin{figure*}
 \begin{center}
   \includegraphics[height=11cm,width=16cm]{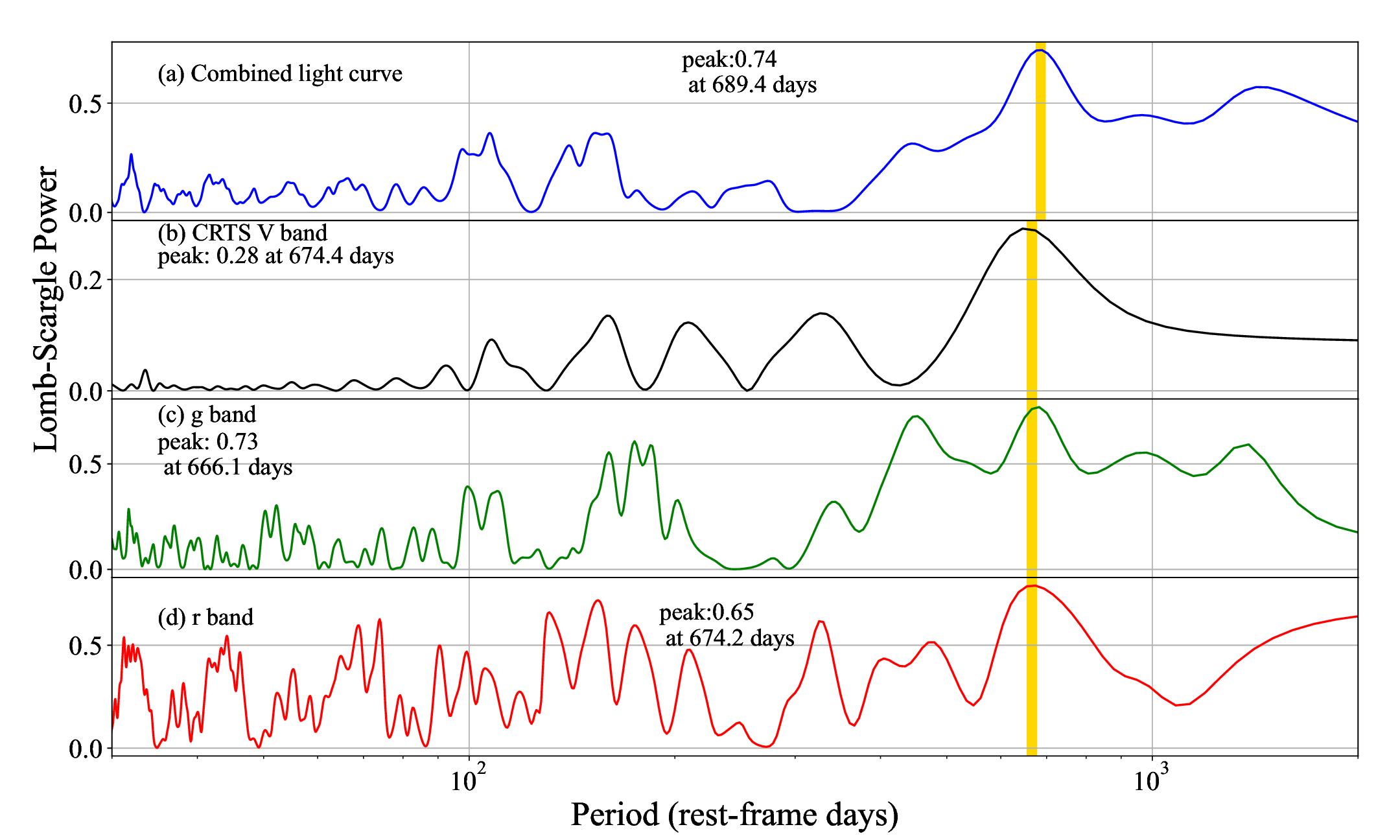} 
 \end{center}
\caption{The Lomb-Scargle periodogram of (a) the combined light curve, (b) CRTS $V$-, (c)
$g$-, and (d) $r$-band light curves including the PanSTARRS, ZTF, and our observation data. 
The yellow vertical lines present the rest-frame day of their power peaks.}
\end{figure*}

\subsection{Auto-correlation analysis}
We performed an auto-correlation function (ACF) analysis, which evaluates 
the correlation between a signal and its delayed replica, so as to verify the 
periodicity of the combined light curve. Since the combined light curve is unevenly 
sampled time-series data, we carried out the $z$-transformed discrete correlation 
function (ZDCF; \cite{1997SpA}) analysis by using ${\tt pyzdcf}$\footnote{$\langle$https://github.com/LSST-sersag/pyzdcf$\rangle$}; 
this method is applicable to uneven sampling data. The ACF shall not be applied to 
three-band light curves that are sparser than the combined light curve.

If a signal is periodic, its ACF is known to decay with periodicity \citep{1993PhR...234..175J}.
We applied an exponentially decaying cosine model to the ACF:
\begin{eqnarray}
{\rm ACF}(\tau)=A\exp(-\lambda \tau)\cos(\omega \tau),
\end{eqnarray}
where $A$, $\lambda$, $\tau$, and $\omega$ are the amplitude, decay rate, time 
lag, and angular frequency, respectively (see also \cite{2020MNRAS.499.2245C}). 
Figure 5 shows the ACF with the ZDCF method. In the combined light curve, the 
error rate between the ACF period ($2\pi/\omega=659.6$ day) and the period 
estimated from the sinusoidal model in table 2 (or the Lomb-Scargle periodogram: 
689.4 day) is $1.0\%$ (or $4.3\%$); the ACF analysis exhibited the periodicity, 
which is comparable to the previous sections.

\begin{figure}
 \begin{center}
   \includegraphics[height=4.5cm,width=8cm]{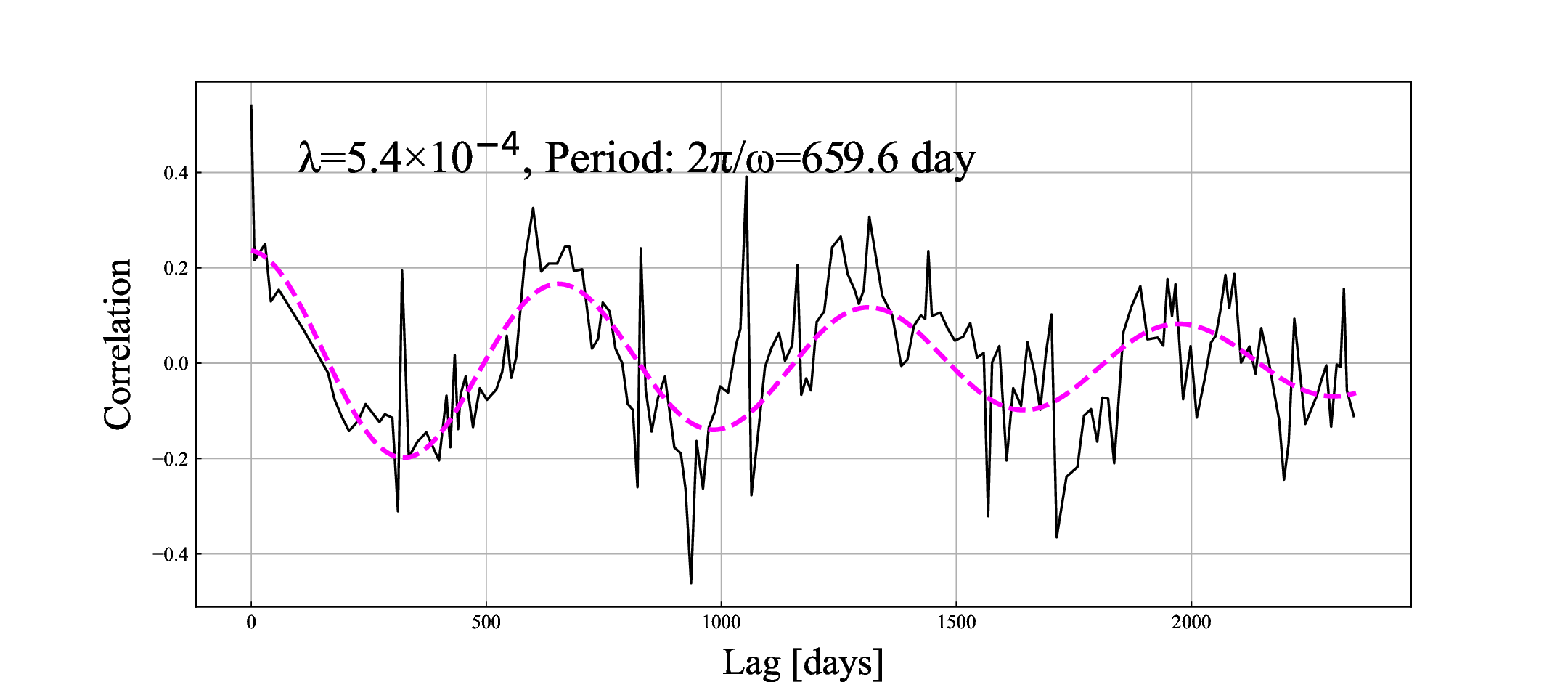} 
 \end{center}
\caption{The ACF of the combined light curve (black solid line) and the best-fit 
decayed cosine model (magenta dashed line) with the ZDCF method.}
\end{figure}

\subsection{The DRW parameters and structure functions}
While the periodicity of the three-band light curves determined by the Lomb-Scargle 
method is consistent with each other, the possibility with which they indicate false-positive 
QPOs must be excluded. The DRW process, which is well used to model AGN 
light curves is known as an aperiodic stochastic process (e.g., \cite{2009ApJ...698..895K};
\cite{2010ApJ...708..927K}; \cite{2010ApJ...721.1014M}). If the light curves of our target 
(figures 2 and 3) are false-positive QPOs, they would behave similarly to the DRW 
or other processes. Here we examine whether the three-band light curves present 
the DRW trend. It should be noted that the DRW is not necessarily a fundamental 
process underlying AGN variability \citep{2016MNRAS.459.2787K}.

The DRW model are well described by the first-order continuous autoregressive [CAR(1)]
process. In this process, the components of the covariance matrix $S$ is expressed as: 
\begin{eqnarray}
S_{ij}=\frac{\tau\sigma^2}{2}\exp\biggr({-\frac{|t_i-t_j|}{\tau}}\biggr),
\end{eqnarray}
where $t_i-t_j$, $\tau$, and $\sigma$ are a time separation, damping timescale, and 
variability amplitude, respectively. In order to evaluate the parameters $\sigma$ and 
$\tau$, we utilized the JAVELIN\footnote{$\langle$https://github.com/nye17/javelin$\rangle$} 
(Just Another Vehicle for Estimating Lags In Nuclei; \cite{2011ApJ...735...80Z}) code, 
which employs the Markov Chain Monte Carlo (MCMC; \cite{2013PASP..125..306F}) 
method for determining posterior distributions of time lags from the reverberation mapping 
and physical parameters, assuming the DRW process. 

In addition to the above analysis, we estimated the structure functions (e.g., 
\cite{1996ApJ...463..466D}; \cite{2004ApJ...601..692V}; \cite{2005AJ....129..615D}; 
\cite{2008MNRAS.383.1232W}; \cite{2010ApJ...721.1014M}; \cite{2018ApJ...866...74S}; 
\cite{2022A&A...664A.117D}) for the three-band light curves expressed as: 
\begin{eqnarray}
SF(\Delta t)=\sqrt{\frac{1}{N(\Delta t)}\sum_{i<j}(\Delta m_{ij})^2},
\end{eqnarray}
where $N(\Delta t)$ is the number of data points in a time-separation bin 
$\Delta t~[=|t_i-t_j|/(1+z)]$, and $\Delta m_{ij}$ is the magnitude difference between 
two epochs. We adopted the structure function definition by \citet{2005AJ....129..615D}, 
which can be evaluated even if the photometric uncertainties are larger than the mean variability 
amplitude. If the behavior of $SF$ follows (or does not follow) the DRW process, 
it would present asymptotic (or jiggly) distribution with increasing time. 
Alternatively, the light curves of WISE J0909+0002 are difficult to express by the DRW 
process, if the asymptotic values of structure functions $SF_{\infty}=\sigma\sqrt{\tau}$ 
significantly deviate from that of normal quasars (see \cite{2010ApJ...721.1014M}).

Figure 6 displays the two-dimension posterior distributions of $\sigma$ and $\tau$ and
the structure functions of the combined light curve and three-band light curves. Using 
those parameters, we estimated the asymptotic value, $SF_{\infty}$, and compared 
our results with those values of figure 3 in \citet{2010ApJ...721.1014M}. Table 3 summarizes 
$\sigma$, $\tau$, and $SF_{\infty}$. The asymptotic values $\log SF_{\infty}$ of 
the three-band light curves significantly deviate from those listed in \citet{2010ApJ...721.1014M}: 
$\log SF_{\infty}\sim-0.70$. While $\log$$SF_{\infty}$ of the combined light curve 
($\log SF_{\infty}\sim-0.57$) was consistent with that of radio-loud quasars in \citet{2010ApJ...721.1014M}, 
the $\tau$-$\sigma$ distribution was away from their results (figure 6a).
\citet{2010ApJ...721.1014M} also calculated $SF_{\infty}$ for quasars with significant 
periodicity (see figure 18 of this reference); the $SF_{\infty}$ values in this study were 
comparable to that study. Moreover, the structure functions exhibited the jiggly distribution 
against the trend of the DRW process (figure 6e). At least in the above analysis, the 
combined light curve and three-band light curves unlikely showed the false-positive QPOs.

\begin{figure*}
 \begin{center}
   \includegraphics[height=10.5cm,width=16cm]{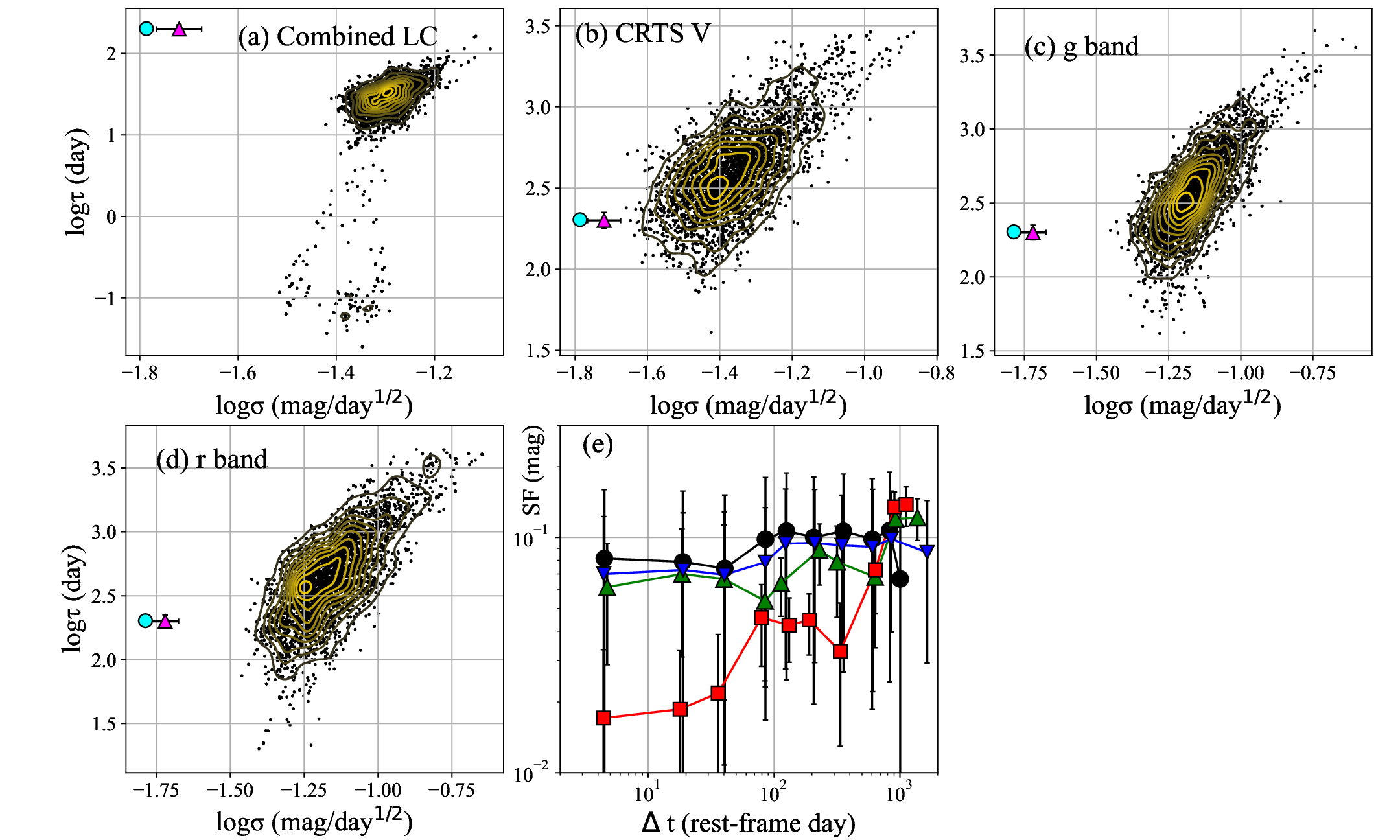} 
 \end{center}
\caption{The two-dimension posterior distributions of variability amplitude $\sigma$ 
and damping timescale $\tau$ for (a) combined light curve, (b) CRTS $V$-,  (c) $g$-,
and (d) $r$-band light curves. The brighter (or dimmer) contours present denser (or thinner) 
regions of these distributions. The cyan circle and magenta triangle indicate no-radio and 
radio-loudest quasar subsamples, respectively (table 2 in \cite{2010ApJ...721.1014M}). 
Panel (e) shows the structure functions of the combined light curve and the three-band 
light curves (inverse triangle: combined light curve, filled circles: CRTS $V$ band, triangles: 
$g$ band, and squares: $r$ band).}
\end{figure*}

\begin{table*}
\tbl{The variability amplitude, damping timescale, and $SF_{\infty}$ of WISE J0909+0002.}{%
\begin{tabular}{cccc}
\hline\noalign{\vskip3pt}
 & $\sigma$ & $\tau$ & $\log SF_{\infty}$\footnotemark[$*$]  \\ 
 & (mag/day$^{1/2}$) & (day) & (mag)  \\ 
\hline
Combined light curve & 0.050$^{+0.005}_{-0.005}$ & 28.74$^{+13.24}_{-11.94}$ & $-$0.571$^{+0.123}_{-0.163}$ \\ 
CRTS $V$ & 0.043$^{+0.013}_{-0.008}$ & 420.97$^{+424.01}_{-193.54}$ & $-$0.054$^{+0.158}_{-0.223}$ \\ 
$g$ & 0.068$^{+0.019}_{-0.014}$ & 354.96$^{+331.50}_{-155.72}$ & 0.107$^{+0.250}_{-0.226}$  \\
$r$ & 0.067$^{+0.026}_{-0.014}$  & 435.22$^{+546.31}_{-234.49}$  & 0.145$^{+0.319}_{-0.270}$ \\
No radio\footnotemark[$\dagger$] & 0.0163 $\pm$ 0.0002 & 201.83 $\pm$ 2.32 & $-$0.634 $\pm$ 0.003 \\
Radio loud\footnotemark[$\dagger$] & 0.019 $\pm$ 0.002 & 199.52 $\pm$ 22.97 & $-$0.57 $\pm$ 0.02 \\
\hline\noalign{\vskip3pt} 
\end{tabular}}\label{table:extramath}
\begin{tabnote}
\hangindent6pt\noindent
\hbox to6pt{\footnotemark[$*$]\hss}\unskip
The asymptotic value of a structure function equal to $\sigma\sqrt{\tau}$. \\
\hbox to6pt{\footnotemark[$\dagger$]\hss}\unskip
Non-radio and radio-loudest quasar subsamples from table 2 in \citet{2010ApJ...721.1014M}.
\end{tabnote}
\end{table*}

\subsection{Simulations of light curves with single power law and damped random walk}
Pure red noise such as the single power-law (SPL) and DRW processes can produce 
quasi-periodic light curves over a few cycles \citep{2016MNRAS.461.3145V}. We generated 
mock light curves with the same baseline as the combine light curve by the SPL and DRW 
simulations, using $\tt{astroML}$\footnote{$\langle$https://www.astroml.org/index.html$\rangle$}, 
so as to compare the sinusoidal model with the pure red noise. As shown in table 4, 
we adjusted the power-law index, $\alpha$, of the SPL model in the range from $-0.5$ to $-$2.0.
The DRW parameters, $\tau$ and $SF_{\infty}$, were set, referring to the results in table 3. 
Since simulated-DRW curves with $\tau\lesssim$ 100 day exhibited extremely large amplitude 
(e.g., 10 mag), we restricted the $\tau$ range from 200 to 600 day. The 10000 light curves were 
generated for each SPL or DRW parameter combination in table 4.

In order to compare the suitability to the sinusoidal model (black solid line in figure 3) among 
the combined light curve, the SPL, and DRW model curves, we evaluated the Bayesian 
information criterion (BIC\footnote{The BIC is defined as, $-2\ln(\mathcal{L})+k\ln(N)$, where 
$\mathcal{L}$, $k$, and $N$ are the likelihood function, the number of free parameters, and the 
number of data, respectively. On the basis of the gaussian error model, the first term is described 
as: $N\ln({\rm RSS}/N)$, where RSS is the residual sum of squares.}) under the gaussian error 
model:  
\begin{eqnarray}
{\rm BIC} = N\ln\biggr\{\frac{\sum_{i=1}^N(y_i - f_{\rm i, sinusoldal})^2}{N}\biggr\}+k\ln(N),
\end{eqnarray}
where $N$ is the number of data, $y_i$ is the i-th observed or simulated data, $f_{\rm i, sinusoldal}$
is the i-th model flux, $k$ is the number of free parameters. We estimated the BIC differences 
$\Delta$BIC$_{\rm Data-SPL}~(=$ BIC$_{\rm sin,data}-$BIC$_{\rm sin,SPL}$) and 
$\Delta$BIC$_{\rm Data-DRW}~(=$ BIC$_{\rm sin,data}-$BIC$_{\rm sin,DRW}$), where
BIC$_{\rm sin,data}$, BIC$_{\rm sin,SPL}$, and BIC$_{\rm sin,DRW}$ are 
the BIC between the best-fit sinusoidal model and the combined light curve, mock SPL 
and DRW curves, respectively. If $\Delta$BIC is less than $-10$ (e.g., \cite{2020MNRAS.499.2245C}; 
\cite{2021MNRAS.500.4025L}), the combined light curve is more suitable with the sinusoidal model 
than mock-light curves by the SPL or DRW simulations. 

As a result, $\Delta$BIC was below $-10$ in most cases. It was difficult to reproduce the light 
curve close to the best-fit sinusoidal model with the SPL simulations. However in the DRW 
simulations with $(\tau,~SF_{\infty})=(400,~0.1)$ and $(600,~0.1)$, $\Delta$BIC$_{\rm Data-DRW}$ 
exceeded $-10$ on rare occasions with a probability of 0.02\% [for $(400,~0.1)$] and 0.08\%
[for $(600,~0.1)$]. Figure 7 displays the best-fit sinusoidal curve and the mock SPL and 
DRW curves, which showed the maximum $\Delta$BIC in the 10000 simulations. For instance, 
the DRW curve in figure 7e presented a sinusoidal-like feature. Therefore, it is difficult to completely 
deny that the light curve of WISE J0909+0002 is the pure red noise in our simulations. More long-term 
monitoring observations would bring us more robust validations for QPOs in WISE J0909+0002. 

\begin{figure*}
 \begin{center}
   \includegraphics[height=10.5cm,width=16cm]{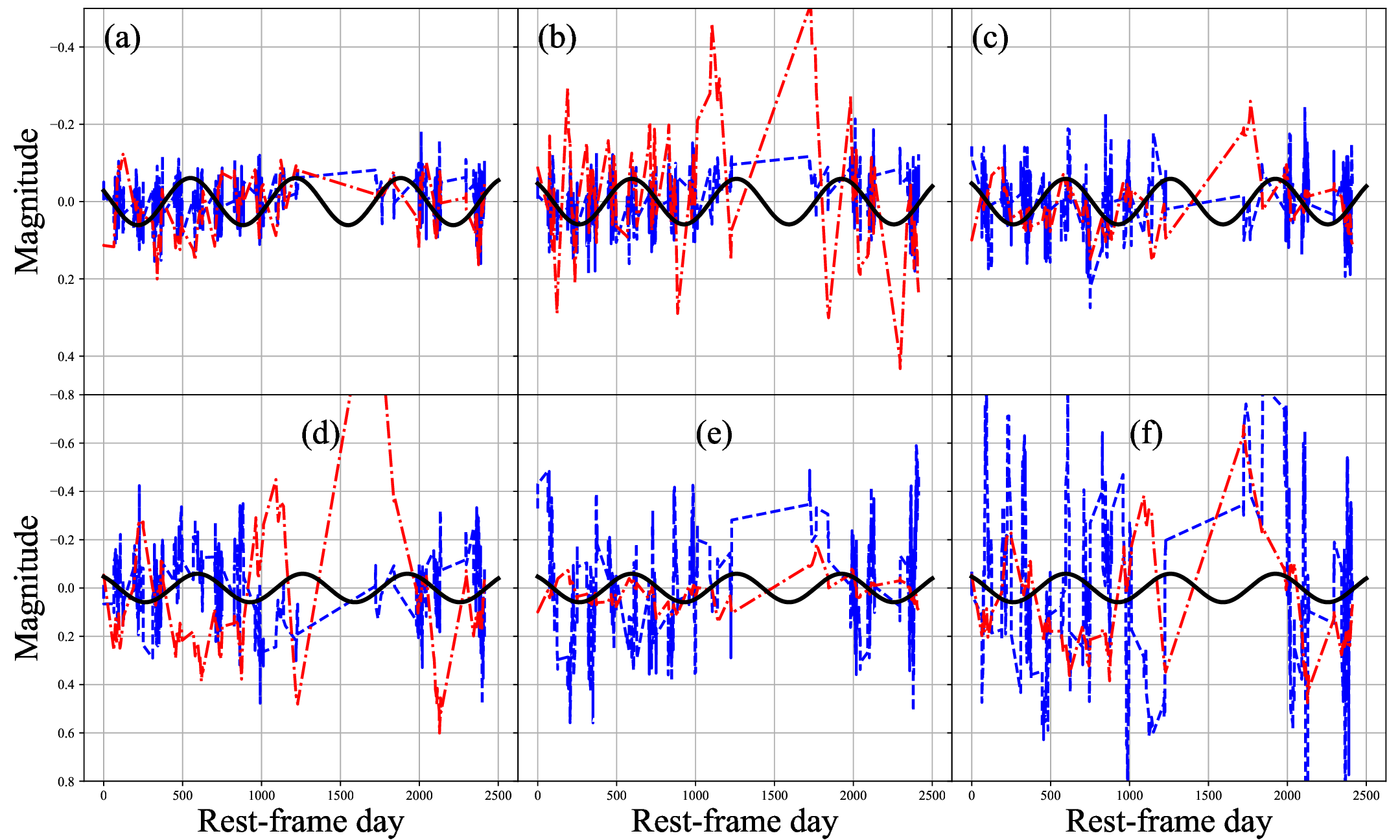} 
 \end{center}
\caption{Mock light curves with the SPL (blue dashed lines) and DRW models (red 
dash-dotted lines), showing the maximum $\Delta$BIC in the 10000 simulations. The black 
solid line is the mean-subtracted (best-fit) sinusoidal model for the combined light curve. 
See table 4 for the SPL and DRW parameters corresponding to each figure.}
\end{figure*}

\begin{table*}
\tbl{The parameters of the SPL and DRW processes and the model comparison with 
the sinusoidal model and red noise.}{
\begin{tabular}{ccccccc}
\hline\noalign{\vskip3pt}
Figure number &  $\alpha$\footnotemark[$*$] & $\Delta$BIC$_{\rm Data-SPL}$\footnotemark[$\dagger$] & $\tau$ & $SF_{\infty}$ & $\Delta$BIC$_{\rm Data-DRW}$\footnotemark[$\ddagger$] \\
& & & (day) & (mag) &  \\ 
\hline
figure 7a & $-$0.5 & $-$138.15 & 200 & 0.1 & $-$293.33 \\ 
7b & $-$0.7 & $-$274.63 & 200 & 0.2 &  $-$1234.78 \\ 
7c & $-$1.0 & $-$581.73 & 400 & 0.1 & 101.72 \\ 
7d & $-$1.2 & $-$1476.40 & 400 & 0.2 & $-$2080.08 \\ 
7e & $-$1.5 & $-$1910.87 & 600 & 0.1 & 313.03  \\ 
7f & $-$2.0 & $-$2615.59 & 600 & 0.2 & $-$1747.82 \\ 
\hline\noalign{\vskip3pt} 
\end{tabular}}\label{table:extramath}
\begin{tabnote}
\hangindent6pt\noindent
\hbox to6pt{\footnotemark[$*$]\hss}\unskip
Power-law index.\\
\hbox to6pt{\footnotemark[$\dagger$]\hss}\unskip
The maximum value of BIC$_{\rm sin,data}-$BIC$_{\rm sin,SPL}$ in the 10000 simulations.\\
\hbox to6pt{\footnotemark[$\ddagger$]\hss}\unskip
The maximum value of BIC$_{\rm sin,data}-$BIC$_{\rm sin,DRW}$ in the 10000 simulations.\\
\end{tabnote}
\end{table*}

\subsection{Power spectral density analysis}
We confirm the periodicity of the combined light curve by the power spectral 
density (PSD) approach. The PSD is useful tool to determine whether signals 
are noisy or periodic. We calculated the PSD, $P(f)$, for the combined light curve 
according to the following equation: 
\begin{eqnarray}
P(f_i) = \frac{2\Delta \tau}{N} |F(f_i)|^2 
\end{eqnarray}
and
\begin{eqnarray}
|F(f_i)|^2=\biggr | \sum^N_{j=1}x_j\cos(2\pi f_i t_j) \biggr |^2 + \biggr | \sum^N_{j=1} x_j\sin(2\pi f_i t_j) \biggr |^2,  
\end{eqnarray}
where $N$, $\Delta t=(t_N-t_1)/N$, $x_j$, and $f_i=i/(N\Delta t)$ are
the number of data, the rest-frame time separation, the mean-subtracted 
flux at time $t_j$, and $i-$th frequency with $i=1, 2,~...,N/2$, respectively 
(e.g., \cite{2002MNRAS.332..231U}; \cite{2019ApJS..241...33L}). 
For modeling the PSD, we considered the (i) SPL (aperiodic)
\begin{eqnarray}
P_{\rm SPL}(f) = Af^{-\alpha},
\end{eqnarray}
(ii) DRW (aperiodic) 
\begin{eqnarray}
P_{\rm DRW}(f) = \frac{A}{1+(f/f_{\rm b})^2},
\end{eqnarray}
and (iii) periodic with aperiodic models
\begin{eqnarray}
P_{\rm P+AP}(f) = \frac{A_{\rm p}}{\sqrt{2\pi}\omega_{\rm p}}\exp\biggr[{-\frac{(f-f_{\rm p})^2}{2\omega_{\rm p}^2}}\biggr]+P_{\rm AP},
\end{eqnarray}
where $A$, $\alpha$, $f_{\rm b}$, $A_{\rm p}$, $f_{\rm p}$, and $\omega_{\rm p}$ are 
free parameters, $P_{\rm AP}$ is $P_{\rm SPL}(f)$ or $P_{\rm DRW}(f)$.
For model selection, we calculated the BIC difference, $\Delta$BIC, with respect 
to the SPL model. 

Figure 8 presents the PSD of the combined light curve with the SPL, DRW, and 
those with periodic model. Although the PSD distribution presented a peak at 802.2 
day, the sparse sampling of the combined light curve resulted in a deviation from
the period obtained from subsections 3.1--3.3 (i.e., 660 to 689 day).  
Table 5 lists the free parameters of the PSD models. The BIC difference, $\Delta$BIC, 
between the SPL and the other models indicates that the SPL$+$Periodic model is 
most preferable of the above models.

\begin{figure*}
 \begin{center}
   \includegraphics[height=10.5cm,width=16cm]{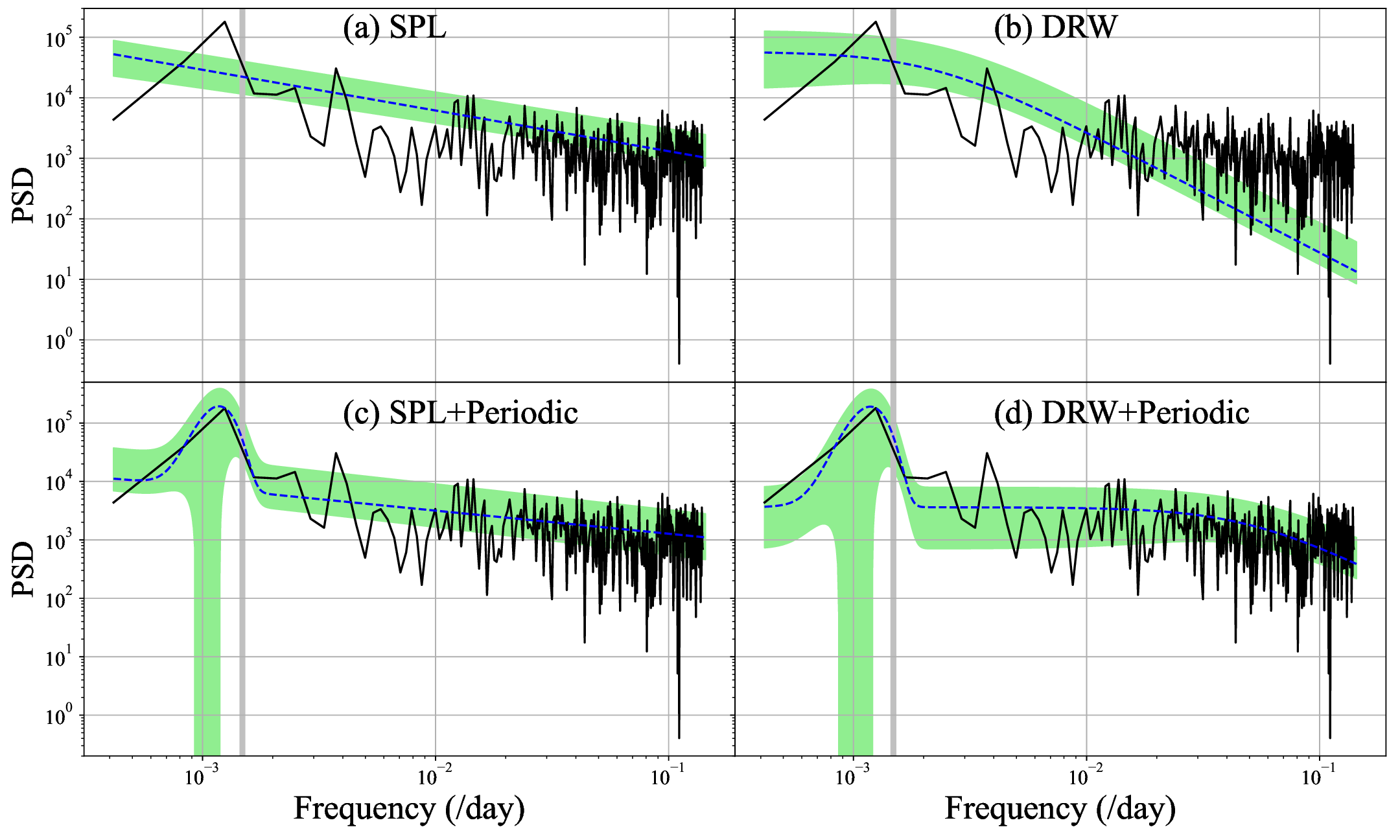} 
 \end{center}
\caption{The PSDs of the combined light curve (black solid lines). Blue dashed lines 
present the best-fit lines for (a) the SPL, (b) DRW, (c) SPL$+$periodic, and (d) 
DRW$+$periodic model. Green shadowed areas indicate the 1$\sigma$ error regions 
for each model. Vertical shadowed regions show the frequency corresponding to the 
660- to 689-day period obtained in subsections 3.1--3.3.}
\end{figure*}

\begin{table*}
\tbl{The model parameters of the PSD.}{%
\begin{tabular}{cccccccc}
\hline\noalign{\vskip3pt}
Model & $\log A$ & $\alpha$ & $\log f_{\rm b}$ & $\log A_{\rm p}$ & $\log f_{\rm p}$ & $\log \omega_{\rm p}$ & $\Delta$BIC\footnotemark[$*$] \\
& & & (day$^{-1}$)  &  & (day$^{-1}$) & (day$^{-1}$) &  \\ 
\hline
SPL & 2.44 $\pm$ 0.17 & $-$0.67 $\pm$ 0.06 & --- & --- & --- & --- & 0 \\ 
DRW & 4.76 $\pm$ 0.05 & --- & $-$2.66 $\pm$ 0.07 & --- & --- & --- & $-$46.04 \\ 
SPL$+$Periodic & 2.71 $\pm$ 0.07 & $-$0.40 $\pm$ 0.04 & --- & 1.93 $\pm$ 0.01 & $-$2.930 $\pm$ 0.004 & $-$3.74 $\pm$ 0.02 & $-$945.62 \\ 
DRW$+$Periodic & 3.56 $\pm$ 0.04 & --- & $-$1.30 $\pm$ 0.06 &1.95 $\pm$ 0.01 & $-$2.928 $\pm$ 0.003 & $-$3.716 $\pm$ 0.003 & $-$917.11 \\ 
\hline\noalign{\vskip3pt} 
\end{tabular}}\label{table:extramath}
\begin{tabnote}
\hangindent6pt\noindent
\hbox to6pt{\footnotemark[$*$]\hss}\unskip
The BIC difference relative to the SPL model.\\
\end{tabnote}
\end{table*}

\section{Discussion}
We found that the light curves of WISE J0909+0002 were likely to show multi-year 
QPOs in the previous section. The period of the combined light curves ($\sim$ 
660$-$689 day in the rest frame) is somewhat longer compared with periodic 
quasars listed in table 2 of \citet{2016MNRAS.463.2145C}. After that, we discuss 
the validity of the following scenarios as mechanisms of QPOs: the Doppler boost, 
circumbinary disk, and radio-jet precession.

\subsection{Doppler boost model}
\subsubsection{Variability amplitude ratio}
Here, we examine the relativistic Doppler boost model qualitatively by using 
multicolor light curves (e.g., \cite{2020MNRAS.499.2245C}; \cite{2021MNRAS.500.4025L}) 
taken with the CRTS, Pan-STARRS, and the OISTER project. In the Doppler 
boost model, photon frequencies are modulated by the factor, 
$D= [\gamma(1-\beta_{\parallel})]^{-1}$, where $\gamma=({1-\beta^2})^{-1/2}$, 
$\beta=v/c$, and $\beta_{\parallel}$ are the Lorentz factor, the three-dimensional 
velocity $v$ in the units of the speed of light $c$, and velocity along our line of sight, 
respectively. The velocity component, $\beta_{\parallel}$, is equal to 
$\beta\cos{\varphi}\sin{i}$, where $\varphi$ is the orbital phase, and $i$ is the orbital 
inclination. The spacial photon density is the Lorentz invariant and is proportional to 
$F_{\nu}/\nu^3$, where $F_{\nu}$ is the apparent flux as a function of the frequency $\nu$. 
Namely, the flux $F_{\nu}$ is expressed as: $F_{\nu}=D^{3-\alpha_{\nu}}F^0_{\nu}$. 
Assuming that intrinsic flux $F^0_{\nu}$ obeys the power-law 
($F^0_{\nu}\propto\nu^{\alpha_\nu}$), the variability amplitude to first order in $\beta$ is 
described as follows: 
\begin{eqnarray}
\frac{\Delta F_{\nu}}{F_{\nu}}=(3-\alpha_{\nu})\beta\cos{\varphi}\sin{i}.
\end{eqnarray}
For instance, the variability amplitude in the CRTS $V$ band is 0.064 mag, which 
corresponds to $\Delta F_{\nu}/F_{\nu}=0.064$. We adopt the phase 
$\cos{\varphi}=1$, since our observation timescale is far shorter than the 
coalescence timescale of BSBHs. 
From equation (11), the ratio of variability amplitude between two bands is written 
as follows:
\begin{eqnarray}
\frac{A_s}{A_l}=\frac{3-\alpha_{\nu, s}}{3-\alpha_{\nu, l}},
\end{eqnarray}
where $A$ is the variability amplitude of sinusoidal curve model. The symbol, 
$s$ (or $l$), is a value for the shorter (or longer) wavelength. 

Using the SDSS DR17 \citep{2022ApJS..259...35A}  spectrum data, we 
estimated the power-law index in each band. The SDSS performed 
two-epoch spectroscopic observations on MJD 51929 and MJD 55532 
for WISE J0909+0002. We estimated the power-law slopes of those SDSS 
spectra to verify the Doppler boost scenario by assuming constant variability 
amplitude in each band thorough out the observation epochs. Table 6 
summarizes the power-law slopes in each band. Figure 9 displays the SDSS 
spectrum of WISE J0909+0002 and the ratios of variability amplitude and 
power-law slopes in each band combination. 

Three of six cases in figure 9b ($g/r$ and $V/r$) and 9d ($g/r$)  
were consistent with the theoretical prediction of the Doppler boost model 
within a 1$\sigma$ accuracy. Thus the scenario is somewhat positive from 
the aspect of the variability amplitude ratios. 

According to the results of \citet{2018MNRAS.476.4617C}, only $\sim$ 20$\%$ 
(or $\sim$ 37$\%$) quasars with periodic light curves showed the Doppler 
boost signature in the NUV (or FUV) region. Integrating the results of 
\citet{2020MNRAS.499.2245C} and \citet{2021MNRAS.500.4025L}, one out of 
five periodic quasars indicated the evidence supporting the Doppler boost scenario 
by the variability amplitude ratios. Thus, we found that WISE J0909+0002 
is likely a rare QPO sample, showing the Doppler boost trend. However, whether the 
Doppler boost theory is supportive depends on the observation wavelength as 
reported in the Swift observations for PG 1302-102 \citep{2020MNRAS.496.1683X}: 
additional X-ray and/or MIR (e.g., \cite{2015...814..L12J}) observations would be 
needed for detailed verifications of this scenario. 


\subsubsection{Parameter space of inclination angle and blackhole mass}
We besides examine the Doppler boost model by exploring parameter spaces 
of the blackhole mass $M_{\rm BH}$, binary blackhole mass ratio $q~=M_2/M_1~(\leq 1)$, 
and orbital inclination $i$ (cf. \cite{2015Natur.525..351D}; \cite{2020MNRAS.499.2245C}; 
\cite{2021MNRAS.500.4025L}; \cite{2021A&A...645A..15S}). In a binary blackhole, the 
velocity of a secondary disk is written as:
\begin{eqnarray}
v_2=\Bigl(\frac{1}{1+q}\Bigr)\Bigl(\frac{2\pi GM_{\rm BH}}{P}\Bigr)^{1/3},
\end{eqnarray}
where $G$ is the gravitational constant, $P$ is the orbital period of a BSBH system.
Using equations (11) and (13) with $\cos{\varphi}=1$ and a primary-blackhole velocity 
$v_1=-qv_2$, the inclination angle $i$ can be expressed as: 
\begin{eqnarray}
i&=& \arcsin{\Bigl\{(1+q)\Bigl(\frac{2\pi GM_{\rm BH}}{P}\Bigr)^{-1/3}\frac{c}{3-\alpha_\nu}\frac{\Delta F_{\nu}}{F_{\nu}}\frac{1}{f}\Bigr\}},
\end{eqnarray}
and
\begin{eqnarray}
f&=& f_2-q(1-f_2), 
\end{eqnarray}
where $f_2$ is the ratio of luminosity contribution from the secondary blackhole. 
We selected the following mass ratios: $q=0$ (extreme case), 0.11, 0.43 
(the characteristic values with two power peaks of mass accretion rates in a 
hydrodynamical simulations; \cite{2014ApJ...783..134F}), and 1.0 (extreme case). 
Figure 10 presents the $M_{\rm BH}$, $i$, and $q$ parameter space of the Doppler 
boost theory for each band (see also \cite{2020MNRAS.499.2245C}, 
\cite{2021MNRAS.500.4025L}). The inclination $i=90^\circ$ (or $i=0$) indicates 
an edge-on (or a face-on) view to accretion disks. Under the conditions of the 
blackhole mass of $7.4\times10^9M_{\odot}$ and the mass-ratio range of 
$0.11\leq q \leq0.43$ at $f_2=0.8$, the inclination angle of WISE J0909+0002 
allows, $i \sim 10^\circ$ (i.e., near the face-on angle).

WISE J0909+0002 is very unusual since it is not only an ELIRG with likely periodic 
variability, but also contains a broad absorption line (BAL; FWHM$~\geq~$2000 
km s$^{-1}$, \cite{1991ApJ...373...23W}) quasar \citep{2017MNRAS.468.4539M}, 
which is thought to eject outflow gas from a viewing angle closer to the edge-on of 
the accretion disk (e.g., \cite{2004ApJ...616..688P}; \cite{2013PASJ...65...40N}). 
Assuming that the BAL quasar fraction ($\sim 10\%-40\%$; \cite{1991ApJ...373...23W}; 
\cite{2003AJ....126.2594R}; \cite{2009ApJ...692..758G}; \cite{2011MNRAS.410..860A}) 
reflects the outflowing angle from the edge-on view (i.e., $i\gtrsim~90^{\circ}-40^{\circ}=
50^{\circ}$), our result ($i \sim 10^\circ$) have a range of considerably smaller angles 
than the above physical picture of BAL quasars. Basing on our analysis for the 
relativistic boost hypothesis the inclination range of WISE J0909+0002 permits the 
claim by which outflows bringing BALs can be ejected from intermediate or close to 
face-on angles \citep{2017MNRAS.467.2571M}. Namely, the physical picture 
of inclination angles based on the detection rate of BALs is inconsistent if 
$f_2>0.8$. WISE J0909+0002 probably ejected outflows near face-on angle 
in the process of evolution to be a BAL quasar (e.g., \cite{2007ApJ...662L..59F}; 
\cite{2009MNRAS.392.1295L}; \cite{2023ApJ...952...44B}) and/or an ELIRG under 
a high $f_2$ ratio.

\begin{table}
\tbl{The spectral power-law slopes of WISE J0909+0002.}{%
\begin{tabular}{ccccc}
\hline\noalign{\vskip3pt}
 & $g$ band & CRTS $V$ & $r$ band  \\ 
\hline
\multicolumn{4}{c}{MJD 55532} \\ 
\hline\noalign{\vskip3pt} 
$\alpha_{\rm \nu}$\footnotemark[$*$] & $-0.558$  $\pm$ 0.027 & $-0.502$ $\pm$ 0.011 & $-0.966$ $\pm$ 0.017 \\
\hline
 \multicolumn{4}{c}{MJD 51929} \\
\hline
$\alpha_{\rm \nu}$\footnotemark[$*$] & $-0.764$  $\pm$ 0.016 & $-0.583$ $\pm$ 0.013 & $-0.111$ $\pm$ 0.021 \\
 \hline\noalign{\vskip3pt} 
\end{tabular}}\label{table:extramath}
\begin{tabnote}
\hangindent6pt\noindent
\hbox to6pt{\footnotemark[$*$]\hss}\unskip
Power-law index of flux as a function of frequency ($=-\beta_{\rm \lambda}-2$), 
where $\beta_{\lambda}$ is the power of the flux as a function of wavelength.\\
\end{tabnote}
\end{table}

\begin{figure*}
 \begin{center}
   \includegraphics[height=11cm,width=16cm]{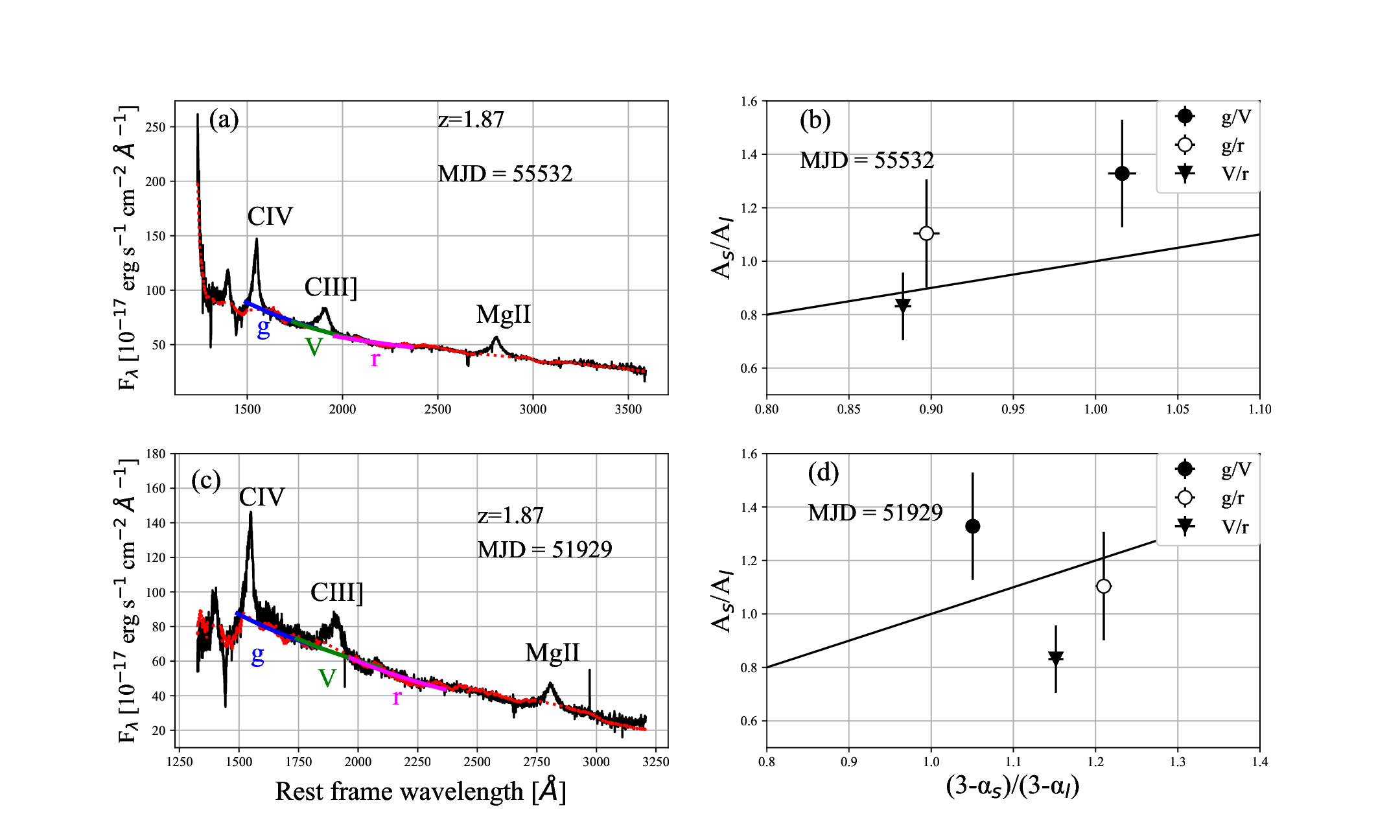} 
 \end{center}
\caption{Panel (a) and (c): the SDSS spectrum of WISE J0909+0002 
(black solid line). Red dotted line presents the model spectrum without 
emission lines. Blue, green, and magenta lines indicate the fitting 
curves to the model spectrum for the $g$-, CRTS $V$-, $r$-band ranges, 
respectively. Characteristic emission lines are labeled. Panel (b) and 
(d): the relation between the variability amplitude ratio and the power-law 
index ratio from equation (12). The theoretical line based on the Doppler 
boost scenario is shown with the black solid line.} 
\end{figure*}

\begin{figure}
 \begin{center}
   \includegraphics[height=14cm,width=8cm]{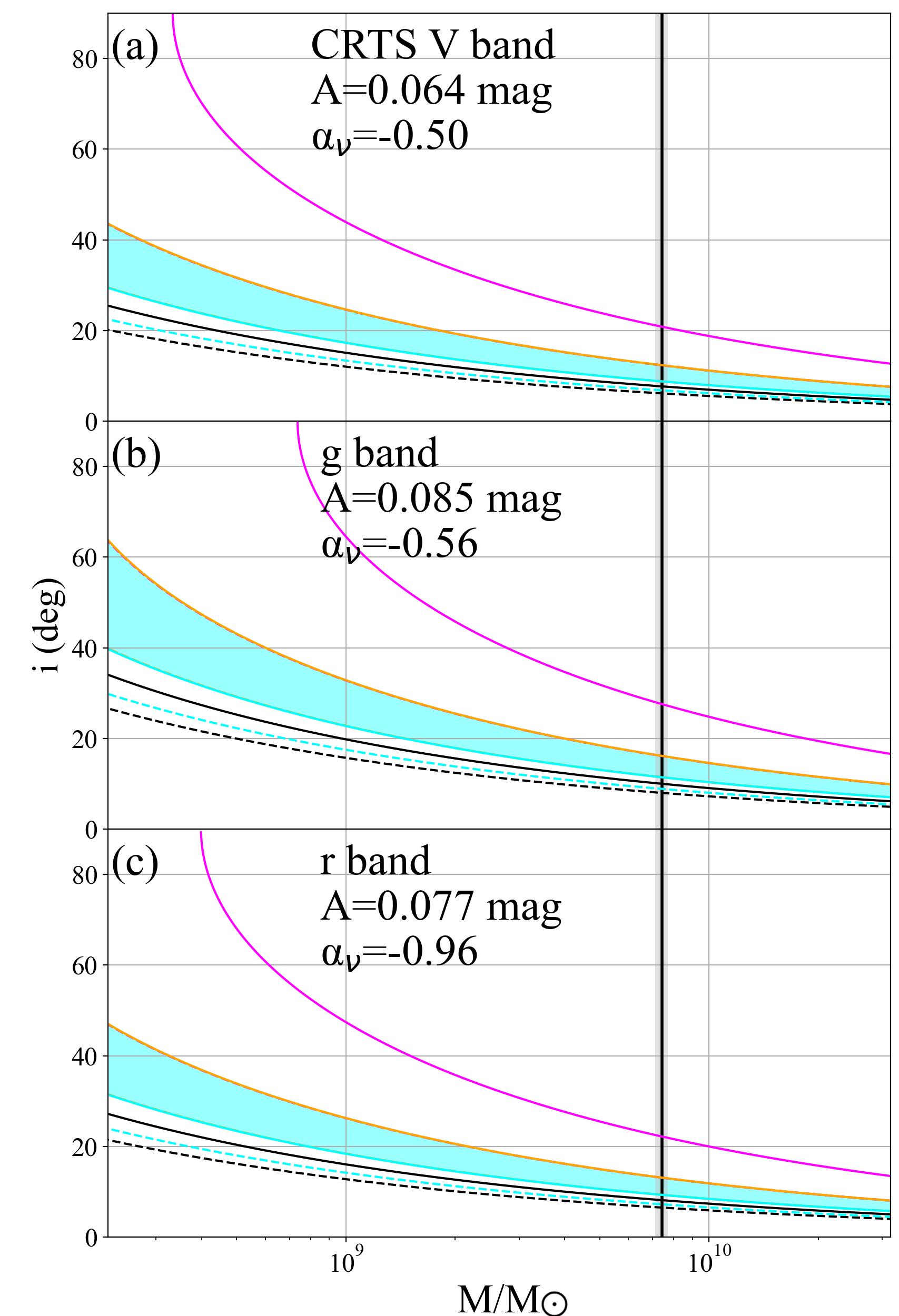} 
 \end{center}
\caption{The parameter space of the blackhole mass, orbital inclination, and 
blackhole mass ratio for the (a) CRTS $V$, (b) $g$, and (c) $r$ bands. 
The black, cyan, orange and magenta curves correspond to mass 
ratios of $q=0,~0.11,~0.43,$ and 1.0, respectively. The dashed (or solid) curves 
show the case where the secondary-disk luminosity contribution, $f_2$, is 1.0 (or 0.8). 
Cyan-shadowed regions indicate the ranges of the inclination angle such that the 
ranges of the mass ratio become $0.11\leq q \leq0.43$ at $f_2=0.8$. The vertical 
solid line and gray shadows represent the blackhole mass of WISE J0909+0002 
(= 7.4 $\times$ 10$^{9}M_{\odot}$) and its 1$\sigma$ error range, respectively.} 
\end{figure}

\subsection{Circumbinary disk model}
As explained in the previous studies, the circumbinary accretion disk (CBD) 
model is one of the plausible candidate of periodic variability in AGNs. This model 
consists of the radiation from the CBD, primary, and secondary minidisks and 
predicts a UV-optical-IR spectral cut-off, which indicates gapped or  
truncated binary disk structures (\cite{2014ApJ...785..115R}; \cite{2015MNRAS.447L..80F}). 
In \citet{2020MNRAS.492.2910G}, six out of the 138 candidates of periodic 
quasars identified by \citet{2015MNRAS.453.1562G} and \citet{2016MNRAS.463.2145C} 
exhibited the spectral cut-offs predicted by the CBD model in the UV-optical-IR 
SEDs. The cut-off temperature is described as: 
\begin{eqnarray}
T_{\rm cut} \sim 2.0\times10^4 \Bigl[\dot{m}\Bigl(\frac{\eta}{0.1}\Bigl)^{-1}\Bigl(\frac{M_{\rm BH}}{10^{8}M_{\odot}}\Bigl)^{-1}\Bigl(\frac{a}{10^2R_{\rm g}}\Bigl)^{-3}\Bigl]^{1/4},
\end{eqnarray}
where $\dot{m}$, $\eta$, $a$, and $R_{\rm g}$ $(= GM_{\rm BH}/c^2)$ are the 
accretion rate in Eddington units, radiative efficiency, the semi-major axis of a binary 
blackhole, and the gravitational radius, respectively. The minimum value of 
a notch appears around $T_{\rm notch}$, where $T_{\rm notch}\sim2^{3/4}T_{\rm cut}$.
Table 7 lists physical parameters for considering the CBD model to the 
spectral energy distributions (SEDs) of our target. 

We obtained the SED (in units of mJy) of WISE J0909+0002 \citep{2021A&A...649L..11T}  
and the composite spectra of 259 type1 quasars \citep{2006ApJS..166..470R}. 
The galactic extinction is corrected for the SED by 
\citet{2021A&A...649L..11T}. Here, we employed the $K$-corrected SED to convert 
the flux into luminosity. Figure 11 displays the normalized SED of WISE J0909+0002. 
According to equations (6), (7), and (8) of \citet{2014ApJ...785..115R}, 
we applied the CBD model to the infrared to FUV flux. From figure 11a, the SED of 
WISE J0909+0002 exhibited a different trend from the CBD model for both 
$q=0.11$ and 0.43 \citep{2014ApJ...783..134F} at $f_2=0.8$: the cut-off and notch 
predicted in the CBD model are not shown in our target. This is the similar result that 
the periodic quasars PSO J334.2028+01.4075 and SDSS J025214.67-002813.7 
did not clearly exhibit the CBD feature (Foord et al. 2017, 2022). 

In addition to the above argument, the NUV and FUV SEDs of WISE J0909+0002 
($\sim$ 10$^{15.5}$ Hz flux in figure 11b) have a steep deficit, which plausibly 
originates from a BSBH accretion mode \citep{2015ApJ...809..117Y} or a reddened 
flux \citep{2014ApJ...788..123L} as discussed in Mrk 231. 
\citet{2021MNRAS.500.4025L} found that the quasar SDSS J025214.67-002813.7 
presented $\sim$ 1.2 dex dropping off near $\sim$1400 ${\rm \AA}$ (figure 11b). 
Then \citet{2022ApJ...927....3F} concluded that the $\sim1400~{\rm \AA}$ flux deficit 
of SDSS J025214.67-002813.7 is better interpreted by a reddened radiation from a 
single accretion disk [$A_V=$ 0.17 and $R(V)=2.54$] rather than a CBD model. 
The estimated dust extinction of WISE J0909+0002 is $E(B-V)=0.13$, which can be 
translated into $A_V = 0.4$ by assuming $R_V = 3.1$ \citep{2021A&A...649L..11T}.
This value is consistent with that of SDSS J025214.67-002813.7. However, $A_V=$ 
0.4 is not particularly large and is below the value required for the dust extinction 
of $A_V \sim 7$ mag \citep{2013ApJ...764...15V}. Furthermore, the rest-frame wavelength 
range from the NUV to FUV bands overlaps with the Lyman limit system 
\citep{2017MNRAS.468.4539M}. Hence, the observed deficit of NUV 
and FUV flux densities is likely due to intrinsic and/or IGM absorptions by 
neutral hydrogen.

\subsection{Precession models}
While the Doppler boost scenario (figures 9 and 10) was favored from our 
results, the CBD model were unlikely (figure 11a), since its model curves was 
not suitable for the SED of WISE J0909+0002. It is necessary to verify other 
plausible scenarios for supporting QPOs such as precessions of accretion disks 
and/or radio jets (e.g., \cite{2012MNRAS.423.3083M}; \cite{2013MNRAS.436L.114K}; 
\cite{2014MNRAS.441.1408T}). For the accretion-disk precession scenario, 
\citet{2022MNRAS.516.3650Z} revealed that the sizes of optical ($37R_{\rm G}$) 
and NUV emission regions ($35R_{\rm G}$) in SDSS J132144+033055 are so 
similar with each other: the disk precession scenario is unlikely to be the cause 
of the periodic flux variability. The radio jet precession hypothesis was also ruled out 
by \citet{2021MNRAS.507.4638C}, since deep 6 GHz radio imaging with NSF’s Karl G. 
Jansky Very Large Array (VLA) for three periodic quasars found that their radio flux was 
too weak (i.e., radio-quiet quasars) to affect UV and optical regions. 

The radio flux of WISE J0909+0002 was not detected by VLA 
FIRST\footnote{$\langle$https://third.ucllnl.org/cgi-bin/firstcutout$\rangle$} \citep{2015ApJ...801...26H}. 
In order to estimate the upper limit of the radio loudness ($R \equiv f_{\rm 6}/f_{\rm 2500}$, 
where $f_{\rm 6}$ and $f_{\rm 2500}$ are the rest-frame flux densities at 6 cm and 2500 ${\rm \AA}$, 
respectively) with a 1 mJy detection threshold, we converted the radio flux limit at 1.4 
GHz to that at 5 GHz, using a power-law in a radio-wavelength region, $F_\nu\propto\nu^{-0.7}$
(e.g, \cite{1992ARA&A..30..575C}; \cite{2019ApJS..243...15T}). Here, the SDSS 
spectrum taken in MJD 55532 was adapted for calculating $f_{\rm 2500}$. 
Consequently, the radio loudness is estimated as an upper limit of $\leq$ 0.4, which 
is difficult to affect the UV/optical flux (i.e., radio-quiet quasar). Thus the radio jet 
precession scenario is rejected for WISE J0909+0002. In \citet{2023A&A...677A...1B}, 
the jet precession (or clear evidence of BSBH nature) was disfavored (or not found) 
from the periodic quasar PSO J334.2028+1.4075. They suggested that a misalignment 
between the inner jet and the outer lobes possibly originated from a warped accretion disk. 
Further observations to determine if broad line regions oscillate and/or accretion disks 
precess would bring new insights into QPO mechanisms or the evolution process of 
ELIRGs. 

\begin{table*}
\tbl{The physical properties of WISE J0909+0002.}{%
\begin{tabular}{ccccccccc}
\hline\noalign{\vskip3pt}
$L_{\rm bol}$\footnotemark[$*$] & $M_{\rm BH}$\footnotemark[$\dagger$]   & $\dot{m}$\footnotemark[$\ddagger$] & $\eta$\footnotemark[$\S$] & $a$\footnotemark[$\|$] & $T_{\rm cut}$\footnotemark[$\#$] & $\lambda_{\rm cut}$\footnotemark[$**$] & $T_{\rm notch}$\footnotemark[$\dagger\dagger$] & $\lambda_{\rm notch}$\footnotemark[$\ddagger\ddagger$] \\   
(10$^{47}$ erg s$^{-1}$)& ($10^9~M_{\odot}$) &&&(ld)&(10$^3$ K)&(10$^3$ {\rm \AA})&(10$^4$ K)&(10$^3$ {\rm \AA})\\ 
\hline\noalign{\vskip3pt} 
4.3 $\pm$ 0.6  & 7.4 $\pm$ 0.3 & 0.4 $\pm$ 0.1 &  0.1 & 29.0 $\pm$ 0.4 & 7.2 $\pm$ 0.5 & 4.0 $\pm$ 0.3 & 1.2 $\pm$ 0.1 & 2.4 $\pm$ 0.1\\ 
 \hline\noalign{\vskip3pt} 
\end{tabular}}\label{table:extramath}
\begin{tabnote}
\hangindent6pt\noindent
\hbox to6pt{\footnotemark[$*$]\hss}\unskip
Bolometric luminosity \citep{2021A&A...649L..11T}.\\
\hbox to6pt{\footnotemark[$\dagger$]\hss}\unskip
Blackhole mass \citep{2021A&A...649L..11T}.\\
\hbox to6pt{\footnotemark[$\ddagger$]\hss}\unskip
Eddington ratio \citep{2021A&A...649L..11T}.\\
\hbox to6pt{\footnotemark[$\S$]\hss}\unskip 
Radiative efficiency. \\
\hbox to6pt{\footnotemark[$\|$]\hss}\unskip
Semi-major axis of a binary blackhole estimated from \citet{2021A&A...645A..15S}.\\
\hbox to6pt{\footnotemark[$\#$]\hss}\unskip
Cut-off temperature.\\
\hbox to6pt{\footnotemark[$**$]\hss}\unskip
Cut-off wavelength based on the Wien's law.\\
\hbox to6pt{\footnotemark[$\dagger\dagger$]\hss}\unskip
Notch temperature. \\
\hbox to6pt{\footnotemark[$\ddagger\ddagger$]\hss}\unskip
Notch wavelength based on the Wien's law.
\end{tabnote}
\end{table*}


\begin{figure*}
 \begin{center}
   \includegraphics[height=8cm,width=16cm]{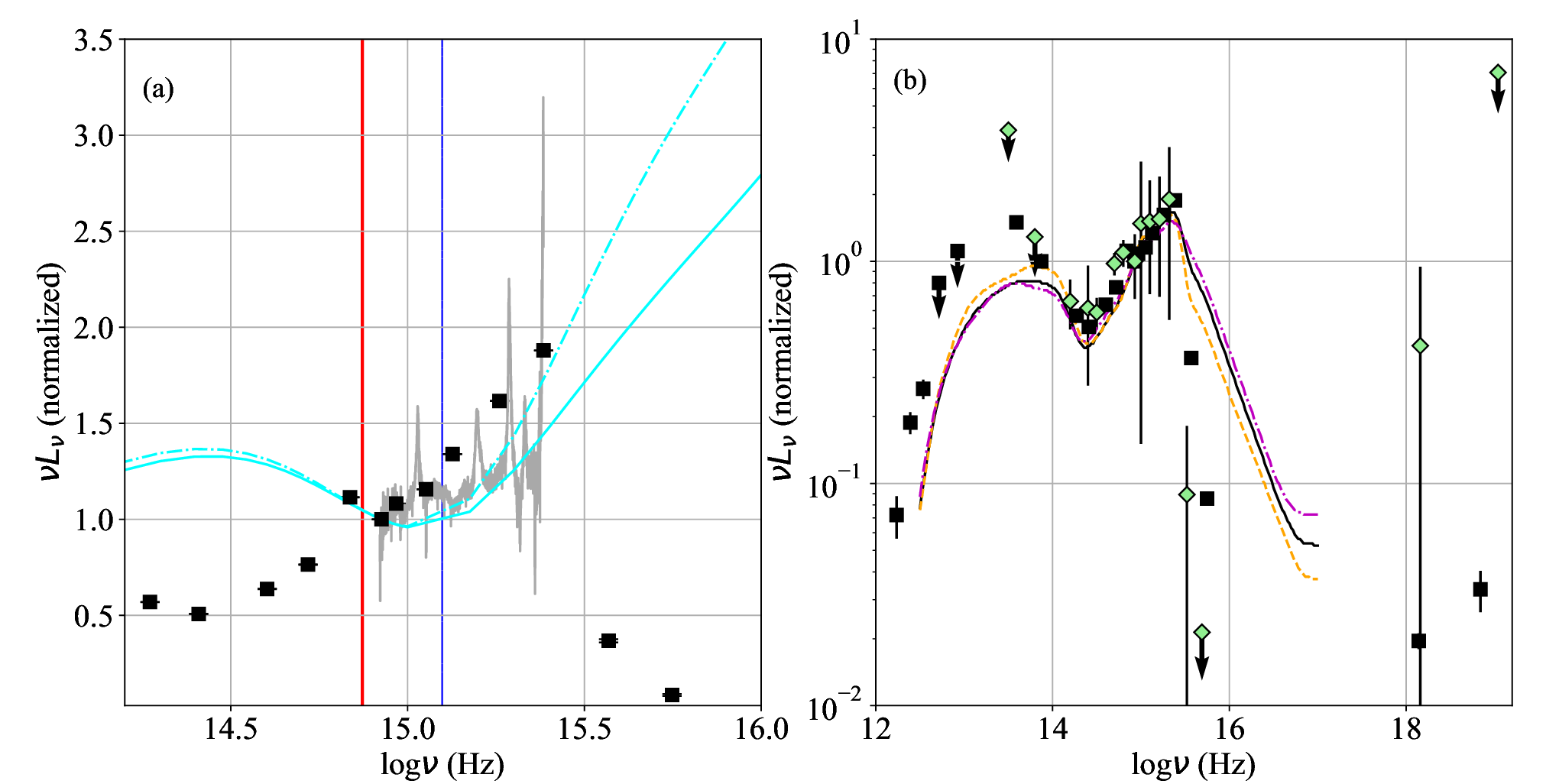} 
 \end{center}
\caption{The rest-frame SED of WISE J0909+0002 (black squares) obtained from 
\citet{2021A&A...649L..11T}. Panel (a): the SED normalized by the rest-frame 3550 
${\rm \AA}$ flux. Gray solid line indicate the rest-frame UV flux obtained from the 
SDSS DR18 catalog. Cyan solid and dash dot lines indicates the CBD model flux with 
mass ratios, $q=0.11$ and 0.43, respectively. The flux ratio of the secondary accretion 
disk, $f_2$, is fixed at 0.8. The vertical red solid (or blue dashed) line shows the positions 
of the cut-off (or notch) wavelength that are expected from the CBD model. Panel (b): 
the flux of WISE J0909+0002, and SDSS J025214.67-002813.7 (green diamonds, 
\cite{2021MNRAS.500.4025L}; \cite{2022ApJ...927....3F}) normalized by each $\sim$ 
10$^{14.93}$ Hz flux. The solid black, orange, and magenta lines denote the composite 
spectra of all quasars (“R06-All”), infrared luminous quasars (“R06-IL”), and infrared 
dim quasars (“R06-ID”), respectively \citep{2006ApJS..166..470R}.} 
\end{figure*}

\section{Conclusion}
WISE J0909+0002 was identified as an ELIRG containing a type-1 quasar 
\citep{2021A&A...649L..11T}. We investigate the periodic flux variability in the 
UV range (quasar-rest frame) by the archival and the observation data 
obtained from the 105 cm Murikabushi telescope, Okayama and Akeno 50 cm 
telescope/MITSuME and the SaCRA 55 cm telescope/MuSaSHI with the 
OISTER collaboration. Our conclusions are summarized as follows: 
\begin{itemize}
	\item[(1)] The CRTS $V$-, $g$-, and $r$-band light curves (figure 2) and the combined 
	light curve (figure 3) show the significant and similar periodicity in each band ($\sim$ 
	660$-$689 day) from the SNR analysis, Lomb-Scargle method (figure 4), and ACF (figure 5);
	 \item[(2)] The light curve periodicity of WISE J0909+0002 plausibly continues at least 
	 $\sim$ 3.6 yr in the rest frame (figure 3);
	  \item[(3)] The $\tau$-$\sigma$ distributions estimated with the JAVELIN 
	  code and the structure functions do not present the DRW trend (figure 6), in other 
	  words it is unlikely that the periodicity of WISE J0909+0002 is a false-positive case 
	  at least in these analyses.
	  \item[(4)] In 10000 simulations, the DRW light curves show periodic-like 
	  curves close to the best-fit sinusoidal model on rare occasions (figure 7), thus the 
	  possibility that the light curves of WISE J0909+0002 are reproduced by the pure red 
	  noise can not completely be rejected.
	  \item[(5)] The PSD of the combined light curve supports the SPL $+$ periodic 
	  model (figure 8).
 	 \item[(6)] The relativistic boost hypothesis is positive from the aspects of 
	 the ratio of variability amplitude and power-law slopes (figure 9) and the existence 
	 of a significant parameter space between the inclination angles and blackhole 
	 mass (figure 10);
   	 \item[(7)] The SED of our target is difficult to be interpreted by the CBD model. 
	 We cannot find cut-offs and/or notches in the SED predicted by this 
	 model (figure 11a). The drop-off of the FUV flux is probably explained by the 
	 intrinsic and/or IGM absorption by neutral hydrogen (figure 11b); 
	 \item[(8)] The radio jet precession scenario is ruled out, since the significant radio 
	 flux density of WISE J0909+0002 is not detected with The VLA FIRST Survey.
\end{itemize}
Our results are generally agree with the relativistic boost scenario. To verify other 
hypothesis of periodic variability such as the warped-disk precession (e.g., 
\cite{2014MNRAS.441.1408T}), follow-up (multi-wavelength) observations are needed. 
In addition, if more longer and more accurate light curves can be obtained, we will be 
able to strengthen the validity of the periodicity and the relativistic boost for 
WISE J0909+0002. 

\section*{Acknowledgments}
We thank the anonymous referee for valuable suggestions. 
This research is (partially) supported by the Optical and Infrared Synergetic 
Telescopes for Education and Research (OISTER) program funded by the MEXT of 
Japan. This work was (partially) carried out by the joint research program of the 
Institute for Cosmic Ray Research (ICRR), The University of Tokyo.
The MITSuME system was supported by a Grant-in-Aid for Scientific Research 
on Priority Areas (19047003). This work was partially supported by JSPS KAKENHI 
Grant Numbers; 23K22537 (YT), 21H01126 and 23K20865 (TM), and 20K14521 (KI). 
TH thank to the staffs in Ishigakijima Astronomical Observatory and Yaeyama 
Star Club for the support in preparing the environment to describe this paper.  

The CSS survey is funded by the National Aeronautics and Space
Administration under Grant No. NNG05GF22G issued through the Science
Mission Directorate Near-Earth Objects Observations Program.  The CRTS
survey is supported by the U.S.~National Science Foundation under
grants AST-0909182 and AST-1313422. MITSuME system was supported 
by a Grant-in-Aid for Scientific Research on Priority Areas (19047003). 

Funding for the Sloan Digital Sky 
Survey IV has been provided by the 
Alfred P. Sloan Foundation, the U.S. 
Department of Energy Office of 
Science, and the Participating 
Institutions. 

SDSS-IV acknowledges support and 
resources from the Center for High 
Performance Computing  at the 
University of Utah. The SDSS 
website is www.sdss.org.

SDSS-IV is managed by the 
Astrophysical Research Consortium 
for the Participating Institutions 
of the SDSS Collaboration including 
the Brazilian Participation Group, 
the Carnegie Institution for Science, 
Carnegie Mellon University, Center for 
Astrophysics | Harvard \& 
Smithsonian, the Chilean Participation 
Group, the French Participation Group, 
Instituto de Astrof\'isica de 
Canarias, The Johns Hopkins 
University, Kavli Institute for the 
Physics and Mathematics of the 
Universe (IPMU) / University of 
Tokyo, the Korean Participation Group, 
Lawrence Berkeley National Laboratory, 
Leibniz Institut f\"ur Astrophysik 
Potsdam (AIP),  Max-Planck-Institut 
f\"ur Astronomie (MPIA Heidelberg), 
Max-Planck-Institut f\"ur 
Astrophysik (MPA Garching), 
Max-Planck-Institut f\"ur 
Extraterrestrische Physik (MPE), 
National Astronomical Observatories of 
China, New Mexico State University, 
New York University, University of 
Notre Dame, Observat\'ario 
Nacional / MCTI, The Ohio State 
University, Pennsylvania State 
University, Shanghai 
Astronomical Observatory, United 
Kingdom Participation Group, 
Universidad Nacional Aut\'onoma 
de M\'exico, University of Arizona, 
University of Colorado Boulder, 
University of Oxford, University of 
Portsmouth, University of Utah, 
University of Virginia, University 
of Washington, University of 
Wisconsin, Vanderbilt University, 
and Yale University.

This study has made use of the NASA/IPAC Infrared 
Science Archive, which is operated by the Jet Propulsion 
Laboratory, California Institute of Technology, under contract 
with the National Aeronautics and Space Administration.

The Pan-STARRS1 Surveys (PS1) and the PS1 public science 
archive have been made possible through contributions by the 
Institute for Astronomy, the University of Hawaii, the Pan-STARRS 
Project Office, the Max-Planck Society and its participating institutes, 
the Max-Planck Institute for Astronomy, Heidelberg and the Max-Planck 
Institute for Extraterrestrial Physics, Garching, The Johns Hopkins 
University, Durham University, the University of Edinburgh, the Queen’s 
University Belfast, the Harvard-Smithsonian Center for Astro- physics, the 
Las Cumbres Observatory Global Telescope Network Incorporated, the 
National Central University of Taiwan, the Space Telescope Science Institute, 
the National Aeronautics and Space Administration under grant No. 
NNX08AR22G issued through the Planetary Science Division of the NASA 
Science Mission Directorate, the National Science Foundation grant No. 
AST-1238877, the University of Maryland, Eotvos Lorand University (ELTE), 
the Los Alamos National Laboratory, and the Gordon and Betty 
Moore Foundation.

Supported by the National Science Foundation under Grants No. AST-1440341 
and AST-2034437 and a collaboration including current partners Caltech, IPAC, 
the Oskar Klein Center at Stockholm University, the University of Maryland, 
University of California, Berkeley, the University of Wisconsin at Milwaukee, 
University of Warwick, Ruhr University, Cornell University, Northwestern University 
and Drexel University. Operations are conducted by COO, IPAC, and UW.

\end{document}